\documentclass[number,preprint,3p]{elsarticle}
\usepackage{graphicx}
\usepackage{lmodern}
\usepackage{amsmath,amssymb,amsthm,mathtools}
\usepackage{booktabs,multirow,float,lineno}
\usepackage{subcaption}
\usepackage{caption}
\usepackage{titlesec}
\titlespacing*{\section}{0pt}{8pt}{2pt}
\titlespacing*{\subsection}{0pt}{6pt}{2pt}
\titlespacing*{\subsubsection}{0pt}{4pt}{2pt}
\usepackage[colorlinks]{hyperref}
\biboptions{numbers,sort&compress,square}
\journal{Reliability Engineering \& System Safety}

\def\tr{^\intercal}
\newcommand{\Prob}[1]{\mathbb{P}\left(#1\right)}
\newcommand{\E}[1]{\mathbb{E}\left[#1\right]}

\newcommand{\cov}[2]{\mathbb{C}ov\left[#1\,,#2\right]}

\newcommand{\vect}[1]{\boldsymbol{#1}}
\newcommand{\say}[1]{``#1''}
\makeatletter
\def\ps@pprintTitle{\let\@oddhead\@empty\let\@evenhead\@empty
  \def\@oddfoot{\hfil\thepage\hfil}\let\@evenfoot\@oddfoot}
\makeatother
\biboptions{numbers,sort&compress,square}

\newcommand{\manuscriptgraphic}[3][]{%
  \IfFileExists{#2}{\includegraphics[#1]{#2}}{%
    \fbox{\begin{minipage}[c][36mm][c]{0.92\linewidth}
      \centering\small Figure asset not supplied.\\[5pt]
      \texttt{\detokenize{#2}}\\[5pt]
      #3
    \end{minipage}}}}

\usepackage{etoolbox}
\allowdisplaybreaks[2]
\makeatletter
\AtBeginDocument{%
    \raggedbottom
    \interlinepenalty=0
    \clubpenalty=50
    \@clubpenalty=50
    \widowpenalty=50
    \displaywidowpenalty=0
    \predisplaypenalty=100
    \postdisplaypenalty=0
    \patchcmd{\@afterheading}
        {\clubpenalty\@M}
        {\clubpenalty\@clubpenalty}
        {}
        {\PackageWarning{layout-tuning}
            {Heading page-break patch failed}}
}
\makeatother

\begin{document}
    \begin{frontmatter}
        \title{Efficient seismic reliability and fragility analysis of lifeline networks \\using subset simulation}
        \tnotetext[published]{Published in \href{https://doi.org/10.1016/j.ress.2025.110947}{Reliability Engineering \& System Safety 260 (2025), 110947}. Received 11 October 2023; revised 27 October 2024; accepted 18 February 2025; available online 19 February 2025.}
        \author[1]{Dongkyu Lee}
        \author[2]{Ziqi Wang\corref{corresponding}}
        \author[3]{Junho Song\corref{corresponding}}
        
        \cortext[corresponding]{Corresponding authors.\ 
            \textit{Email addresses:}
            \href{mailto:ziqiwang@berkeley.edu}
                 {\texttt{ziqiwang@berkeley.edu}} (Ziqi Wang),
            \href{mailto:junhosong@snu.ac.kr}
                 {\texttt{junhosong@snu.ac.kr}} (Junho Song).}
        \address[1]{Engineering Risk Analysis Group, Technical University of Munich, Munich, Germany}
        \address[2]{Department of Civil and Environmental Engineering, University of California, Berkeley, CA, USA}
        \address[3]{Department of Civil and Environmental Engineering, Seoul National University, Seoul, South Korea}
        \begin{abstract}
        Various simulation-based and analytical methods have been developed to evaluate the seismic fragilities of individual structures. However, the seismic safety and resilience of a community are substantially affected by network reliability, determined not only by component fragilities but also by network topology and commodity/information flows. Seismic reliability analyses of networks often encounter significant computational challenges due to complex network topologies, interdependencies among ground motions, and low failure probabilities. This paper proposes a variance-reduction method for network fragility analysis using subset simulation to overcome these challenges. The binary network limit-state function is reformulated into more informative piecewise continuous functions. The proposed limit-state functions quantify the proximity of each sample to the network failure domain, thereby enabling the construction of specialized intermediate failure events, which can be utilized in subset simulation and other sequential Monte Carlo approaches. Moreover, by identifying an implicit relationship between intermediate failure events and seismic intensity, we propose a technique to obtain the entire network fragility curve with a single execution of specialized subset simulation. Numerical examples demonstrate that the proposed method can effectively evaluate system-level fragility for large-scale networks.
        \end{abstract}
    \end{frontmatter}

    \vspace{-12pt}
    \section{Introduction}
    \noindent Lifeline networks, such as transportation, gas, and electricity systems, are the critical backbone of modern society. Their significance is particularly noticeable in the post-hazard stage because emergency assessment, evacuation, life-saving, and repair operations rely on the functionality of lifeline networks. Therefore, it is essential to quantify the network reliability to design and maintain resilient lifeline networks against seismic hazards \cite{troisi2022technical,du2023regional,oh2024long}. To this end, this paper aims to develop an efficient sampling method to assess the seismic reliability and fragility of networks under various earthquake magnitudes.

    To quantify the impact of earthquakes on lifeline networks, various network reliability metrics, e.g., connectivity, flow capacity, and travel time, have been proposed \cite{chen2002capacity,lin2003flow,ching2007efficient,esposito2015simulation}. For example, connectivity reliability, such as two-terminal or $k$-terminal reliability, evaluates network accessibility in terms of the probability that at least one origin-destination (OD) pair remains connected. By contrast, capacity reliability is the probability that the flow capacity between OD pairs surpasses a certain threshold. In transportation networks, this analysis can be combined with traffic demand flow analysis to ascertain the reliability in terms of travel time \cite{asakura1991road, bell1999sensitivity, chen2002capacity}.

    In large-scale networks, however, network reliability analysis can be computationally challenging or even infeasible due to the high computation cost, intricate network topology, or interdependencies between components. To mitigate the computational challenge known as “combinatorial explosion,” various simulation-based approaches \cite{chen2021bridge, liu2021research, hou2022seismic} are extensively used, including the crude/direct Monte Carlo simulation (MCS), owing to their broad applicability and flexibility. However, the crude MCS has a slow convergence rate of $\mathcal{O}(N^{-1/2})$, where $N$ is the number of random sample points. This slow convergence rate may lead to a prohibitive computation cost in rare event simulations, such as the failures of lifeline networks. Furthermore, analyzing the performance of each sample point is time-consuming in large-scale lifeline networks. Surrogate models for network reliability indices \cite{rocco2002fast, stern2017accelerated, nabian2018deep} have been developed to reduce the computation time. However, these surrogate models entail inherent errors, which can be exacerbated in low-probability events.

    Advanced sampling techniques \cite{au2001estimation, kurtz2013cross, wang2016cross, yang2017cross, geyer2019cross, wang2019hamiltonian, chen2022riemannian, xian2024relaxation} can accelerate the probability estimation by sampling in critical regions with higher probabilities. In particular, subset simulation (SS) \cite{au2001estimation} relaxes the target failure event into nested intermediate failure events, effectively estimating low probabilities with relatively small samples and working efficiently even in high-dimensional spaces. An essential ingredient of SS is ranking sample points according to their limit-state function (LSF) values so intermediate failure events can be formulated. However, most network LSFs in network reliability analyses have binary or multi-state outputs. This feature poses significant challenges for SS because the sample points outside the failure domain may have the same LSF value; consequently, the algorithm cannot move toward the failure domain as there is no information to guide the sampling in the correct direction. Ching and Hsu \cite{ching2007efficient} proposed a continuous LSF for network reliability analysis using random walks. However, the computation cost for simulating random walks becomes high in large-scale networks. Chan et al. \cite{chan2022adaptive} suggested adapting the number of sample points and conditional probabilities to avoid sampling the same LSF value, but this “adaptive effort subset simulation (aE-SuS)” approach does not differ from the MCS for binary LSFs. On the other hand, the sequential space conversion (SESC) method \cite{rashki2021sesc}, which combines control variates with space conversion techniques in SS, can be directly applied to binary network LSFs through scaling. However, the convergence from the scaled LSFs to the true LSF inevitably introduces efficiency loss compared with the traditional SS.
    
    The primary contribution of this paper is to propose two piecewise continuous reformulations of the binary LSF representing network disconnection. These new LSFs enable the construction of relaxed, intermediate failure events, readily usable in SS and alternative sequential sampling methods \cite{papaioannou2016sequential, xian2024relaxation}. Since the two reformulations involve trade-offs in accuracy and efficiency, one can select the LSF that aligns better with the analysis goals. Another main contribution of this paper is an alternative interpretation of the intermediate failure domains in the context of subset simulation-based network fragility analysis. An implicit connection between intermediate failure events and earthquake magnitude enables a single run of SS to generate the entire network fragility curve. The proposed method can be readily extended to analyze $k$-terminal reliability and $k$-out-of-$N$ reliability (focusing on $k$-out-of-$N$:$G$, but applicable to $k$-out-of-$N$:$F$), which are computationally challenging for non-simulation-based approaches \cite{barlow1984computing, wu1994algorithm}.
    
    The paper is structured as follows. Section \ref{sec2} provides an overview of the seismic network reliability analysis. Section \ref{sec3} briefly reviews SS, develops informative network LSFs, and proposes a computational framework for network fragility analysis. In Section \ref{sec4}, three numerical examples demonstrate the performance of the proposed method. Finally, Section \ref{sec5} summarizes the paper and provides future research directions. 
    
    \section{Overview of seismic network reliability analysis}\label{sec2}
    \subsection{Ground motion intensity at network components and failure probabilities}\label{sec21}
    \noindent For seismic network reliability analysis, one should first assess the seismic risk of individual components$/$ structures in a network. The seismic failure of a component is defined as the event that the seismic demand exceeds the seismic capacity, both of which are uncertain.

    \subsubsection{Intensity of ground motions}\label{sec211}
    \noindent Various ground-motion intensity measures (IMs), such as peak ground acceleration (PGA), peak ground velocity (PGV), and spectral acceleration ($S_{a}$), are used to quantify seismic demands. One can either adopt one of these IMs or use multiple IMs \cite{baker2005vector, luco2007structure}. In this paper, PGA is used to quantify the intensity of seismic demand, which can be easily extended to $S_a$ by interpolation \cite{chopra1995dynamics, kurtz2016seismic}. The attenuation relation of PGA can be expressed as \cite{abrahamson1992stable, joyner1993methods}
    \begin{equation}\label{Eq1}
    \ln{D_i}=f(M,R_i,\vect{\mathbf{\lambda}}_i)+\eta+\varepsilon_i,
    \end{equation}
    where $D_i$ is the PGA at site $i$; $f(\cdot)$, denoted hereafter as $\ln{\bar{D}_i}$, is the attenuation relation for the PGA at site $i$ as a function of magnitude $M$, the distance from the epicenter to site $i$, $R_i$, and a set of other explanatory variables $\vect{\mathbf{\lambda}}_i$; and $\eta$ and $\varepsilon_i$ are the inter- and intra-event residuals with zero means and standard deviations $\sigma_\eta$ and $\sigma_\varepsilon$, respectively. Since IMs are generally modeled as lognormal random variables, both residuals $\eta$ and $\varepsilon_i$ are assumed to follow normal distributions \cite{goda2008spatial}.

    Since the components in a single network can be close to each other, their IMs and seismic responses can be highly correlated. Specifically, the Pearson correlation coefficient between the PGAs at sites $i$ and $j$ arises from the common variable $\eta$ and the correlation between the intra-event residuals $\varepsilon_i$ and $\varepsilon_j$. It is often assumed that $\eta$ and $\varepsilon_i$ are statistically independent, and the correlation coefficient between $\varepsilon_i$ and $\varepsilon_j$ is given as a function of the distance $\Delta_{ij}$ between the two sites. From Eq.~\eqref{Eq1}, the correlation coefficient between $\ln{D_i}$ and $\ln{D_j}$ is derived as \cite{goda2008spatial}
    \begin{equation}\label{Eq2}
    \rho_{\ln{D_i}\ln{D_j}}(\Delta_{ij})=\frac{\cov{\ln{D_i}}{\ln{D_j}}}{\sigma_{\ln{D_i}}\sigma_{\ln{D_j}}}=\frac{\sigma_\eta^2+\rho_{\varepsilon_i\varepsilon_j}(\Delta_{ij})\sigma_\varepsilon^2}{\sigma_\eta^2+\sigma_\varepsilon^2}.
    \end{equation}

    The examples in this paper adopt the attenuation relation model in \cite{boore2008ground} to predict the mean of the natural logarithm of the PGA demand at the $i^\text{th}$ component as
    \begin{equation}\label{Eq3}
    \ln{\bar{D}}_i=-0.5265-0.0115\sqrt{R_i^2+1.35^2}+\ln{(R_i^2+1.35^2)}[-0.3303+0.0599(M_w-4.5)],
    \end{equation}
    where $M_w$ is the moment magnitude; and both $R_i$ and $\Delta_{ij}$ are given in km. In this paper, the standard deviations of the inter- and intra-event residuals, $\sigma_\eta$ and $\sigma_\varepsilon$, are set to 0.265 and 0.502, respectively \cite{lim2012efficient}. The intra-event spatial correlation is calculated using the model proposed in \cite{goda2008spatial} as
    \begin{equation}\label{Eq4}
    \rho_{\varepsilon_i\varepsilon_j}(\Delta_{ij})=\exp{(-0.27\Delta_{ij}^{0.40})}.
    \end{equation}

    \subsubsection{Component failure probabilities and correlation coefficients}\label{sec212}
    \noindent The seismic capacity of component $i$, $C_i$, is modeled by a lognormal distribution with a median $\bar{C}_i$ and lognormal standard deviation $\zeta_i$ \cite{nielson2007seismic, lim2012efficient, lim2015seismic, lee2021multi}. Then, the Bernoulli variable $B_i$ representing the failure event of component $i$, i.e., the event when the seismic demand $D_i$ exceeds the seismic capacity $C_i$, is defined as
    \begin{equation}\label{Eq5}
    B_i=\mathbb{I}(C_i\leq D_i)=\mathbb{I}(z_i\leq0),
    \end{equation}
    where $\mathbb{I}(\cdot)$ denotes a binary indicator function that returns 1 if the given inequality holds, and 0 otherwise; and $z_i=\ln{C_i}-\ln{D_i}$ denotes the logarithmic safety margin \cite{der2022structural}. The seismic demand $D_i$ and the seismic capacity $C_i$ are assumed to be statistically independent. Since both $D_i$ and $C_i$ follow lognormal distributions, $z_i$ follows a normal distribution with a mean $\mu_{z_i}=\ln{\bar{C}}_i-\ln{\bar{D}}_i$ and a variance $\sigma_{z_i}^2=\zeta_i^2+\sigma_\eta^2+\sigma_\varepsilon^2$. Then, the seismic failure probability of component $i$, $P_i$, is given as
    \begin{equation}\label{Eq6}
    P_i=\E{B_i}=\E{\mathbb{I}(z_i\leq0)}=\Phi(-\beta_i),
    \end{equation}
    where $\Phi(\cdot)$ is the standard normal cumulative distribution function (CDF); and $\beta_i=\mu_{z_i}/\sigma_{z_i}$ denotes the reliability index.

    Eq.~\eqref{Eq6} can be extended to derive the joint failure probability of components numbered from 1 to $N$ as
    \begin{equation}\label{Eq7}
    \Prob{\bigcap_{i=1}^N\{z_i\leq0\}}=\Phi_N(-\vect{\mathbf{\beta}},\vect{\mathbf{R}}_{\vect{\mathbf{zz}}}),
    \end{equation}
    where $\Phi_N(\cdot,\cdot)$ is the $N$-variate zero-mean, unit-variance normal CDF; $\vect{\mathbf{\beta}}=[\beta_1,\ldots,\beta_N]\tr$ is the vector of reliability indices; $\mathbf{R}_\mathbf{zz}=[\rho_{z_iz_j}]_{i,j\in[1,N]}$ is the $N\times N$ correlation matrix, which is equivalent to the covariance matrix in the present context; and $\rho_{z_iz_j}$ is the correlation coefficient between $z_i$ and $z_j$. Lee and Song \cite{lee2021multi} analytically derived the correlation coefficient $\rho_{z_iz_j}$ from Eq.~\eqref{Eq2} as
    \begin{equation}\label{Eq8}
    \rho_{z_iz_j}=\frac{\zeta_i\zeta_j\delta_{ij}+\sigma_\eta^2+\sigma_\varepsilon^2\rho_{\varepsilon_i\varepsilon_j}(\Delta_{ij})}{\sqrt{\zeta_i^2+\sigma_\eta^2+\sigma_\varepsilon^2}\sqrt{\zeta_j^2+\sigma_\eta^2+\sigma_\varepsilon^2}},
    \end{equation}
    where $\delta_{ij}$ is the Kronecker delta, which is 1 when $i=j$, and 0 otherwise. Eq.~\eqref{Eq8} dramatically reduces the time required to compute the correlation coefficients compared to numerical methods while maintaining accuracy.

    If the seismic capacities do not follow the lognormal distributions as assumed in this paper, one can use the first- and second-order approximations \cite{der2022structural} of the reliability index $\beta_i$ and Eq.~\eqref{Eq8}, i.e.,
    \begin{gather}
    \beta_i\cong\frac{\ln{\mu_{C_i}}-0.5\delta_{C_i}^2-\ln{\bar{D}}_i}{\sqrt{\delta_{C_i}^2+\sigma_\eta^2+\sigma_\varepsilon^2}},
    \label{Eq9}\\
    \rho_{z_iz_j}\cong\frac{\delta_{C_i}\delta_{C_j}\delta_{ij}+\sigma_\eta^2+\sigma_\varepsilon^2\rho_{\varepsilon_i\varepsilon_j}(\Delta_{ij})}{\sqrt{\delta_{C_i}^2+\sigma_\eta^2+\sigma_\varepsilon^2}\sqrt{\delta_{C_j}^2+\sigma_\eta^2+\sigma_\varepsilon^2}},
    \label{Eq10}
    \end{gather}
    where $\mu_{C_i}$, $\sigma_{C_i}$, and $\delta_{C_i}=\sigma_{C_i}/\mu_{C_i}$ denote the mean, standard deviation, and coefficient of variation ($c.o.v.$) of $C_i$, respectively.

    \subsection{Network reliability analysis}\label{sec22}
    \noindent A lifeline network consists of line-type components, such as pipelines and roads, and node-type components, such as stations and bridges. The network can be described by a graph $G=(\vect{V},\vect{E})$, where $\vect{V}$ denotes the set of nodes (or vertices) representing both types of components, and $\vect{E}$ is the set of links (or edges) indicating the conceptual connectivity between nodes. That is, it is assumed that all links in $\vect{E}$ are perfectly reliable. This assumption will not cause error in the network reliability analysis because nodes represent the physical entities. The above assumption still holds for networks with link failures by a polynomial-time conversion to equivalent networks with node failures \cite{colbourn1987network, ball1995network}.

    Consider a network state vector $\vect{\mathbf{z}}=[z_1,\ldots,z_N]$, denoting a vector of the logarithmic safety margins of components, where $N=|\vect{V}|$ is the number of nodes (i.e., the total number of node-type and line-type components) in the network of interest. The network reliability problem computes the network failure probability $P_f$ through an $N$-fold integral in the space of the network state vector, i.e.,
    \begin{equation}\label{Eq11}
    P_f=\int_\mathcal{F}f_{\vect{\mathbf{Z}}}(\vect{\mathbf{z}})d\vect{\mathbf{z}}=\int_{\mathbb{R}^N}\mathbb{I}(G(\vect{\mathbf{z}})\leq0){f_{\vect{\mathbf{Z}}}(\vect{\mathbf{z}})d\vect{\mathbf{z}}},
    \end{equation}
    where $\mathcal{F}=\{G(\vect{\mathbf{z}})\leq0\}$ is the failure domain for the network reliability problem, such as connectivity reliability (e.g., two-terminal reliability) and capacity reliability \cite{zuev2015general}; $G(\vect{\mathbf{z}})\in\mathbb{R}$ is the network limit-state function (LSF); and $f_{\vect{\mathbf{Z}}}(\vect{\mathbf{z}})$ is the joint probability density function (PDF) of the network state vector $\vect{\mathbf{z}}$.

    For the two-terminal reliability between an origin-destination (OD) node pair, the network LSF in Eq.~\eqref{Eq11} is defined as the binary LSF
    \begin{equation}\label{Eq12}
    G_{\text{OD}}^\text{Bi}(\vect{\mathbf{z}})=\left\{\begin{aligned}&1,&&\text{if the OD pair is connected in }\vect{\mathbf{z}},\\&0,&&\text{otherwise},\end{aligned}\right.
    \end{equation}
    which depends on the network topology. For example, in a series system, only the joint survival of all components guarantees connectivity. In contrast, a parallel system will fail if and only if all components fail. That is, the failure domains of $N$-component series and parallel systems, $\mathcal{F}_\text{series}$ and $\mathcal{F}_\text{parallel}$, are defined respectively as
    \begin{gather}
    \mathcal{F}_\text{series}=\{G_\text{series}^\text{Bi}(\vect{\mathbf{z}})=0\}=\bigcup_{i=1}^N\{B_i=1\}=\left\{\min_{i=[1,N]}z_i\leq0\right\},
    \label{Eq13}\\
    \mathcal{F}_\text{parallel}=\{G_\text{parallel}^\text{Bi}(\vect{\mathbf{z}})=0\}=\bigcap_{i=1}^N\{B_i=1\}=\left\{\max_{i=[1,N]}z_i\leq0\right\},
    \label{Eq14}
    \end{gather}
    where $G_\text{series}^\text{Bi}$ and $G_\text{parallel}^\text{Bi}$ respectively denote the cases of binary network LSFs with series and parallel systems for an OD pair.

\begin{figure}[!htbp]
        \centering
        \includegraphics[width=.9\textwidth]{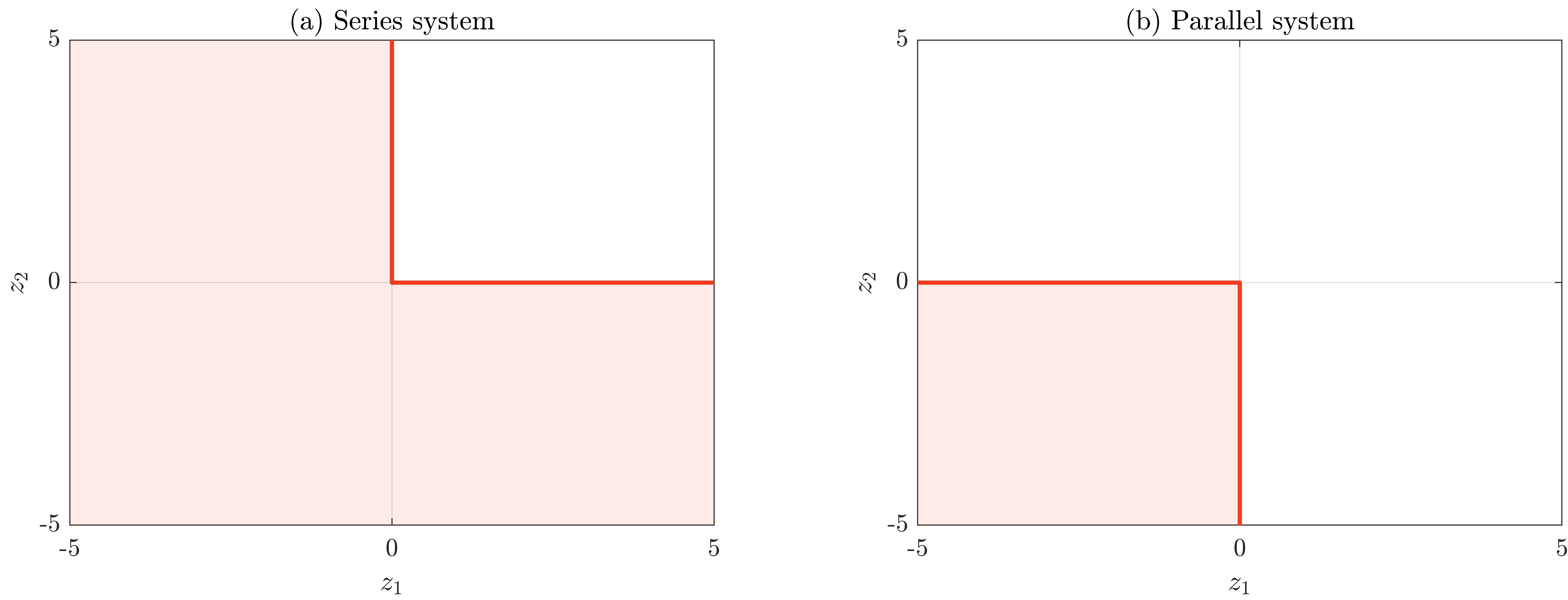}
        \caption{Failure domains of (a) two-component series system; and (b) two-component parallel system.}
        \label{Fig1}
    \end{figure}

    For example, consider two-component series and parallel systems. According to Eqs.~\eqref{Eq13} and \eqref{Eq14}, the failure domains for $\mathcal{F}_\text{series}$ and $\mathcal{F}_\text{parallel}$ are shown in red in Fig.~\ref{Fig1}(a) and (b), respectively. The exact failure probability $P_f$ of each system can be evaluated by integrating $f_{\vect{\mathbf{Z}}}(\vect{\mathbf{z}})$ over each failure domain $\mathcal{F}$.

    Series and parallel systems can be considered as networks with a link set (i.e., a set of nodes forming a single path between the OD pair) and a cut set (i.e., a set of nodes whose simultaneous failure results in the disconnection of the OD pair), respectively. In other words, by generalizing Eqs.~\eqref{Eq13} and \eqref{Eq14}, respectively, the failure domain of a general network, $\mathcal{F}$, is given in two ways \cite{song2003bounds} as
    \begin{gather}
    \mathcal{F}=\bigcap_{\forall k}{\bigcup_{i\in L_k}\{B_i=1\}}=\left\{\max_k{\min_{i\in L_k}z_i}\leq0\right\},
    \label{Eq15}\\
    \mathcal{F}=\bigcup_{\forall l}{\bigcap_{i\in C_l}\{B_i=1\}}=\left\{\min_l{\max_{i\in C_l}z_i}\leq0\right\},
    \label{Eq16}
    \end{gather}
    where $L_k$ and $C_l$ denote the $k^\text{th}$ link set and the $l^\text{th}$ cut set, respectively.

    Once all the cut sets or link sets in a network with $N$ components are identified, the exact network reliability can be evaluated by combining Eq.~\eqref{Eq11} with Eq.~\eqref{Eq15} or \eqref{Eq16}. To this end, several non-simulation-based methods have been developed to identify all link sets \cite{abraham1979improved,aziz1993enumeration} or cut sets \cite{brown1971computerized, rosenthal1979approaches}. In addition, one can use partial link sets and cut sets to compute the upper and lower bounds of the network reliability \cite{li2002recursive, lim2012efficient, lim2015seismic, lee2021multi}. However, reliability problems for general networks are represented by highly nonlinear LSFs, because they involve max and min operators and path searching; under certain conditions, there could even be percolation-style phase transition. That is, network reliability analysis problems are $NP$-hard \cite{rosenthal1977computing, colbourn1987network}, i.e., no polynomial-time algorithm exists. Therefore, non-simulation-based approaches can be inappropriate for large-scale networks. In summary, reliability analyses of large-scale networks often face challenges in (1) exploration of the failure domain in $2^{N}$ component state combinations, and (2) fast and accurate computation of probability in the high-dimensional space $\mathbb{R}^{N}$.
            
    \section{Subset simulation for network reliability analysis}\label{sec3}    
    \subsection{Review of subset simulation}\label{sec31}
    \noindent Subset simulation (SS) \cite{au2001estimation} is one of the most widely used variance-reduction sampling approaches. In SS, the failure domain of interest, $\mathcal{F}$, is represented by $m$ nested intermediate failure domains, $\mathcal{F}_1\supset\mathcal{F}_2\supset\dots\supset\mathcal{F}_m=\mathcal{F}$. The failure probability $P_f$ is expressed as the product of the conditional probabilities $\Prob{\mathcal{F}_k|\mathcal{F}_{k-1}}$, i.e.,
    \begin{equation}\label{Eq17}
    P_f=\prod_{k=1}^m{\Prob{\mathcal{F}_k|\mathcal{F}_{k-1}}},
    \end{equation}
    where $\mathcal{F}_0=\mathbb{R}^N$ denotes the initial null failure domain. By setting each conditional probability (except the last) equal to a constant $p_0$, the intermediate failure domains $\mathcal{F}_k$, $k=1,2,\ldots,m-1$, are adaptively determined by the $p_0$ quantile of LSF values associated with sample points in $\mathcal{F}_{k-1}$. Au and Beck \cite{au2001estimation} proposed setting $p_0=0.1$, and Zuev et al. \cite{zuev2012bayesian} demonstrated that $p_0\in[0.1,0.3]$ has optimal performance.

    While generating independent and identically distributed samples from the initial null failure domain is typically feasible and straightforward, it becomes challenging for the intermediate failure domains $\mathcal{F}_{k-1}$, $k\geq2$. To this end, Markov Chain Monte Carlo (MCMC) methods, such as the Metropolis-Hastings algorithm \cite{metropolis1953equation,hastings1970monte,papaioannou2015mcmc}, can be utilized. Using an MCMC method, each conditional probability $\Prob{\mathcal{F}_k|\mathcal{F}_{k-1}}$, and the failure probability estimate $\hat{P}_{f,SS}$, can be expressed as follows, respectively:
    \begin{gather}
    \Prob{\mathcal{F}_k|\mathcal{F}_{k-1}}=\int_{\mathcal{F}_k}{f(\vect{\mathbf{z}}|\mathcal{F}_{k-1})d\vect{\mathbf{z}}}\cong\frac{1}{n}\sum_{j=1}^n{\mathbb{I}(\vect{\mathbf{z}}^{(j)}\in\mathcal{F}_k|\mathcal{F}_{k-1})},
    \label{Eq18}\\
    \hat{P}_{f,SS}=\prod_{k=1}^m{\Prob{\mathcal{F}_k|\mathcal{F}_{k-1}}}\cong\frac{p_0^{m-1}}{n}\sum_{j=1}^n{\mathbb{I}\left(\vect{\mathbf{z}}^{(j)}\mathcal{\in F}|\mathcal{F}_{m-1}\right)},
    \label{Eq19}
    \end{gather}
    where $n$ is the number of sample points generated in each intermediate failure domain; and $\vect{\mathbf{z}}^{(j)}$ is the $j^\text{th}$ sample point. SS is particularly efficient for rare events because the number of samples required for a single run of SS is $n_\text{SS}\propto|\log{P_f}|$ \cite{au2001estimation}, while the crude Monte Carlo simulation (MCS) requires $n_\text{MCS}\propto 1/P_f$ simulations.

    MCMC methods have a critical impact on the performance of SS; ideally, the MCMC sample should show limited random walk behavior and achieve rapid mixing. In this work, we adopt the Hamiltonian Monte Carlo-based subset simulation (HMC-SS) \cite{wang2019hamiltonian,chen2022riemannian}, an efficient variant of SS leveraging the desirable properties of HMC.

    \subsection{Informative network limit-state function for subset simulation}\label{sec32}
    \noindent The two-terminal reliability is typically represented by the binary LSF in Eq.~\eqref{Eq12}. This property is a major obstacle to using SS in network reliability analysis. For a binary function, the $p_0$ quantile of the samples is chosen to be either 0 or 1 in each intermediate domain, so SS may not identify the failure domain effectively. To address this problem, $G_\text{OD}^\text{Bi}(\vect{\mathbf{z}})$ should be reformulated as a multi-state or continuous function.

    \subsubsection{Most reliable path-based network limit-state function}\label{sec321}
    \noindent We propose an informative network LSF, which encodes the same failure domain as the original binary function but provides additional information on the direction and distance to the failure domain. To this end, we introduce the most reliable path (RP) concept, defined as the path between an OD pair with the highest probability of survival \cite{lim2012efficient}. It follows that the network LSF is defined based on the vulnerability of the RP as follows:
    \begin{equation}\label{Eq20}
    G_{\text{OD}}^\text{RP}(\vect{\mathbf{z}})=\left\{\begin{aligned}&\frac{\min_{i\in\vect{\mathbf{RP}}}z_i}{n_\text{RP}},&&\text{if the OD pair is connected in }\vect{\mathbf{z}},\\&0,&&\text{otherwise},\end{aligned}\right.
    \end{equation}
    where $\vect{\mathbf{RP}}$ denotes the set of nodes on the most reliable path with positive $z_i$; and $n_\text{RP}$ is the number of nodes in $\vect{\mathbf{RP}}$. The LSF proposed in Eq.~\eqref{Eq20} stems from the observation that (1) the network fails if the RP fails, and (2) the larger $n_\text{RP}$ is, the more likely the RP (a series system) is to fail. Because of the denominator $n_\text{RP}$, the proposed function is piecewise rather than globally continuous; there may be a discontinuity along the boundaries where RP changes, which can lead to inefficient sampling and inaccurate estimates.

\begin{figure}[htbp]
        \centering
        \includegraphics[width=.9\textwidth]{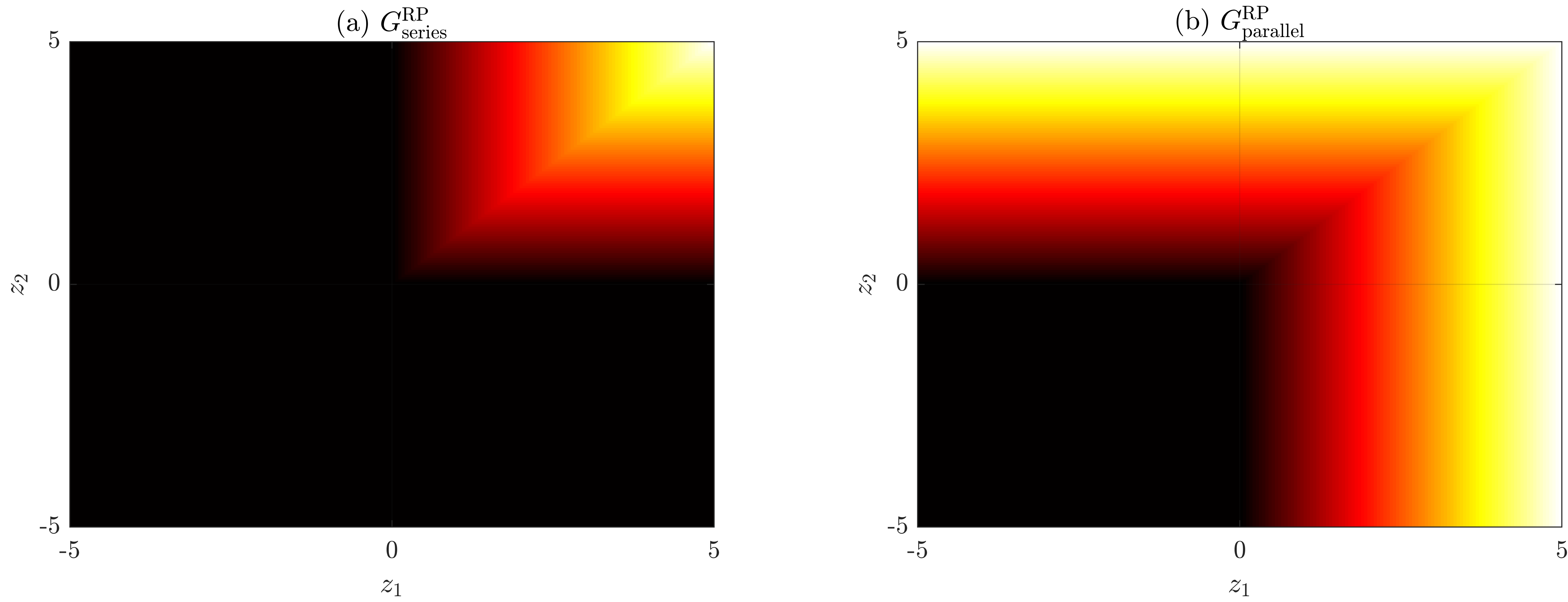}
        \caption{Proposed RP-based network LSF for (a) two-component series system; and (b) two-component parallel system.}
        \label{Fig2}
    \end{figure}

    To find the RP, all component failure events are assumed to be independent. This independence assumption is used only to identify the RP. Then, Dijkstra's algorithm \cite{ahuja1993network, cormen2009introduction} is used to find the RP that maximizes the product of the survival probabilities of nodes obtained from Eq.~\eqref{Eq6}, i.e., the sum of the log-scaled survival probabilities. Fig.~\ref{Fig2}(a) and (b) visualize the proposed network LSFs by color maps (black: disconnection, white: robust connection) for the two-component series and parallel systems, which are contrasted with the binary functions in Fig.~\ref{Fig1}(a) and (b).

    Fig.~\ref{Fig2} shows that the proposed function monotonically decreases as it approaches the failure domain, so the intermediate domain gradually converges to the failure domain as SS progresses. Let the $k^\text{th}$ intermediate failure domain $\mathcal{F}_k$ be $\{G_\text{OD}^\text{RP}(\vect{\mathbf{z}})\leq g_k\}$. Then, according to Eq.~\eqref{Eq19}, $P_f$ is expressed as the product of conditional probabilities as
    \begin{equation}\label{Eq21}
    \begin{aligned}\hat{P}_{f,SS}=&\prod_{k=1}^m{\Prob{G_\text{OD}^\text{RP}(\vect{\mathbf{z}})\leq g_k|G_\text{OD}^\text{RP}(\vect{\mathbf{z}})\leq g_{k-1}}}\\\cong&\frac{p_0^{m-1}}{n}\sum_{j=1}^n{\mathbb{I}\left(G_\text{OD}^\text{RP}\left(\vect{\mathbf{z}}^{(j)}\right)\leq g_m|G_\text{OD}^\text{RP}\left(\vect{\mathbf{z}}^{(j)}\right)\leq g_{m-1}\right)},\end{aligned}
    \end{equation}
    where $g_1>\dots>g_m=0$ denote intermediate thresholds; $g_0=\infty$ denotes the initial failure threshold; and $\vect{\mathbf{z}}^{(j)}$ is the state vector of the components in the $j^\text{th}$ network-state sample point.

    \subsubsection{Shortest path-based network limit-state function}\label{sec322}
\begin{figure}[htbp]
        \centering
        \includegraphics[width=.9\textwidth]{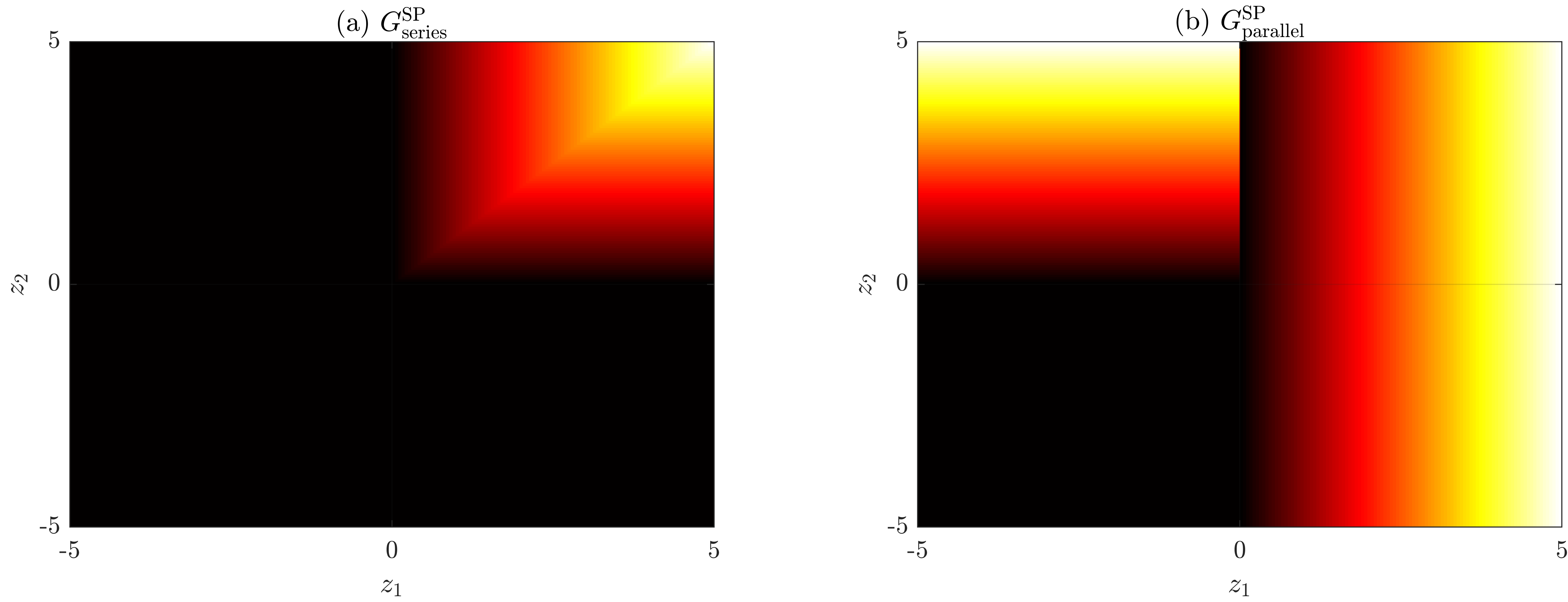}
        \caption{Proposed SP-based network LSF for (a) two-component series system; and (b) two-component parallel system.}
        \label{Fig3}
    \end{figure}

    \noindent The proposed most reliable path-based function $G_\text{OD}^\text{RP}(\vect{\mathbf{z}})$ has a critical flaw in terms of computation time for reliability analysis of large-scale networks; the time complexity of Dijkstra's algorithm for the weighted graphs is given by $\mathcal{O}(|\vect{V}|^2)$ \cite{dijkstra1959note}, which is quite time-consuming compared to the breadth-first search (BFS) for the OD connectivity, i.e., $G_\text{OD}^\text{Bi}(\vect{\mathbf{z}})$, which has a linear time complexity of $\mathcal{O}(|\vect{E}|+|\vect{V}|)$. Although SS requires fewer simulations than the crude MCS, the high computation cost per sample can offset the benefits. To compensate for this weakness, by replacing the RP with the shortest path (SP), another LSF that utilizes the BFS is proposed as
    \begin{equation}\label{Eq22}
    G_{\text{OD}}^\text{SP}(\vect{\mathbf{z}})=\left\{\begin{aligned}&\frac{\min_{i\in\vect{\mathbf{SP}}}z_i}{n_\text{SP}},&&\text{if the OD pair is connected in }\vect{\mathbf{z}},\\&0, &&\text{otherwise},\end{aligned}\right.
    \end{equation}
    where $\mathbf{SP}$ denotes the set of nodes on the SP consisting of nodes with positive $z_i$; and $n_\text{SP}$ is the number of nodes on $\mathbf{SP}$. Fig.~\ref{Fig3}(a) and (b) visualize the proposed SP-based network LSF by color maps (black: disconnection, white: robust connection) for the two-component series and parallel systems, $G_\text{series}^\text{SP}(\vect{\mathbf{z}})$ and $G_\text{parallel}^\text{SP}(\vect{\mathbf{z}})$. Fig.~\ref{Fig3}(a) for $G_\text{series}^\text{SP}(\vect{\mathbf{z}})$ is the same as Fig.~\ref{Fig2}(a) for $G_\text{series}^\text{RP}(\vect{\mathbf{z}})$, since there is only a single path in a series system, i.e., $\mathbf{SP}\equiv\mathbf{RP}$. On the other hand, there is an apparent difference between the two color maps in the parallel system; in contrast to $G_\text{parallel}^\text{RP}(\vect{\mathbf{z}})$ in Fig.~\ref{Fig2}(b), $G_\text{parallel}^\text{SP}(\vect{\mathbf{z}})$ has the discontinuity along the boundary between the first and second quadrants because the SP considers only the first component when both survive (i.e., the first quadrant in Fig.~\ref{Fig3}(b)). This distinctively shaped network LSF results in a larger variance of SS estimates but has a lower computational cost than the RP-based function.

    \subsubsection{Extension to $k$-terminal reliability \& $k$-out-of-$N$ reliability}\label{sec323}
    \noindent The analysis of network capacity reliability is much more complex than that of two-terminal reliability. It requires different methods tailored to the characteristics of each reliability problem (e.g., BFS for connectivity reliability, Ford-Fulkerson algorithm \cite{ford1956maximal} for capacity reliability). In contrast, simulation-based analysis can be used for different network reliability problems in the same way as for the two-terminal reliability problem once their LSFs are well defined.

    For example, we can consider $k$-terminal reliability \cite{nabian2018deep}, a generalization of two-terminal reliability. More specifically, $k$-terminal reliability is defined as the probability that all nodes in $\vect{V}_O$ are connected to all nodes in $\vect{V}_D$, where $\vect{V}_O$ and $\vect{V}_D$ denote the sets of origin and destination nodes, respectively, and $k=|\vect{V}_O|+|\vect{V}_D|$. The network LSFs for two-terminal reliability are extended to the LSF for $k$-terminal reliability, $G_k(\vect{\mathbf{z}})$, as
    \begin{equation}\label{Eq23}
    G_k(\vect{\mathbf{z}})=\min_{\text{O}_i\in\vect{V}_O,\text{D}_j\in\vect{V}_D}{G_{\text{O}_i\text{D}_j}(\vect{\mathbf{z}})}.
    \end{equation}

    Because connectivity is checked repeatedly as many times as the number of OD pairs per sample, the computational complexity of $k$-terminal reliability evaluation grows proportionally with $k$.

    Furthermore, the LSF for $k$-out-of-$N$ reliability, $G_{k/N}$, i.e., the probability that at least $k$ among $N$ OD pairs are connected \cite{byun2017reliability}, can be formulated as
    \begin{equation}\label{Eq24}
    G_{k/N}(\vect{\mathbf{z}})={\text{maxk}_{i\in[1,N]}}\left({G_{\text{OD}_i}(\vect{\mathbf{z}}),k}\right),
    \end{equation}
    where $\text{maxk}(\cdot,j)$ is defined as the function that returns the $j^\text{th}$ largest value. It is noteworthy that $G_{k/N}$ becomes identical to $G_k$, when $k=N$. In analytical methods \cite{barlow1984computing, wu1994algorithm}, the time complexity for $k$-out-of-$N$ reliability is given by $\mathcal{O}(N\cdot k)$. However, when utilizing order statistic functions like Eq.~\eqref{Eq24}, the complexity remains proportional to $N$, independent of $k$.

    \subsection{Framework to assess network seismic fragility curves}\label{sec33}
    \noindent On top of estimating the network failure probability for one earthquake magnitude, the proposed informative network LSFs also enable SS to evaluate network fragility curves. In particular, the intermediate failure domains in SS are now redefined as the failure domain under each $M_w$, and their probabilities correspond to discretized points on a fragility curve, with the x-axis representing the magnitude and the y-axis describing the network failure probability. To this end, this section introduces the process of configuring the intermediate failure domains and generating the network fragility curve.

    \subsubsection{Configuration of the intermediate failure domains}\label{sec331}
\begin{figure}[H]
        \centering
        \includegraphics[width=.68\textwidth]{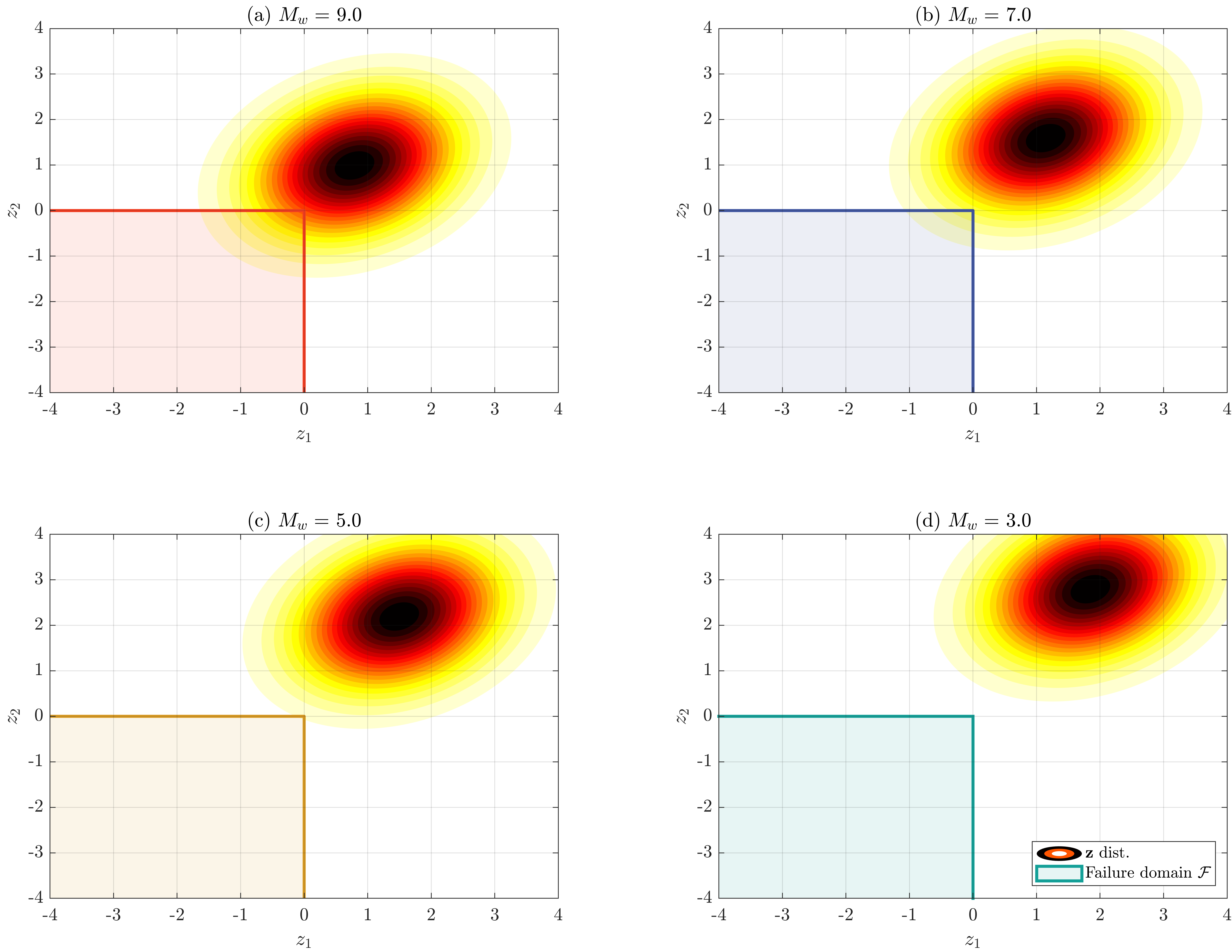}
        \caption{Failure domains of two-component parallel system and color maps of $\vect{\mathbf{z}}$ at different $M_w$.}
        \label{Fig4}
    \end{figure}

    \noindent Unlike individual structures, lifeline networks are distributed over a large area. Since IMs are measured differently across all sites for the same earthquake, it is considered more appropriate to use $M_w$ as the x-axis in the network fragility curves. Consider the case where $M_w$ changes, while the epicenter remains fixed. In that case, $\vect{\mathbf{z}}$ is represented as a function of $M_w$, i.e., $\vect{\mathbf{z}}(M_w)$. While seismic demands depend on $M_w$, seismic capacities as well as the inter- and intra-event residuals remain unaltered regardless of $M_w$. In other words, as $M_w$ varies, the covariance matrix $\mathbf{R}_\mathbf{zz}=[\rho_{z_iz_j}]_{N\times N}$ remains constant, and only the mean of $\vect{\mathbf{z}}(M_w)$ changes. For example, Fig.~\ref{Fig4} shows the isodensity map of $\vect{\mathbf{z}}(M_w)$ at several values of $M_w$, and the failure domains of a two-component parallel system, with more details provided in \ref{A1}. The failure domains are identical to those in Fig.~\ref{Fig1}(b), regardless of $M_w$.

\begin{figure}[htbp]
        \centering
        \includegraphics[width=.38\textwidth]{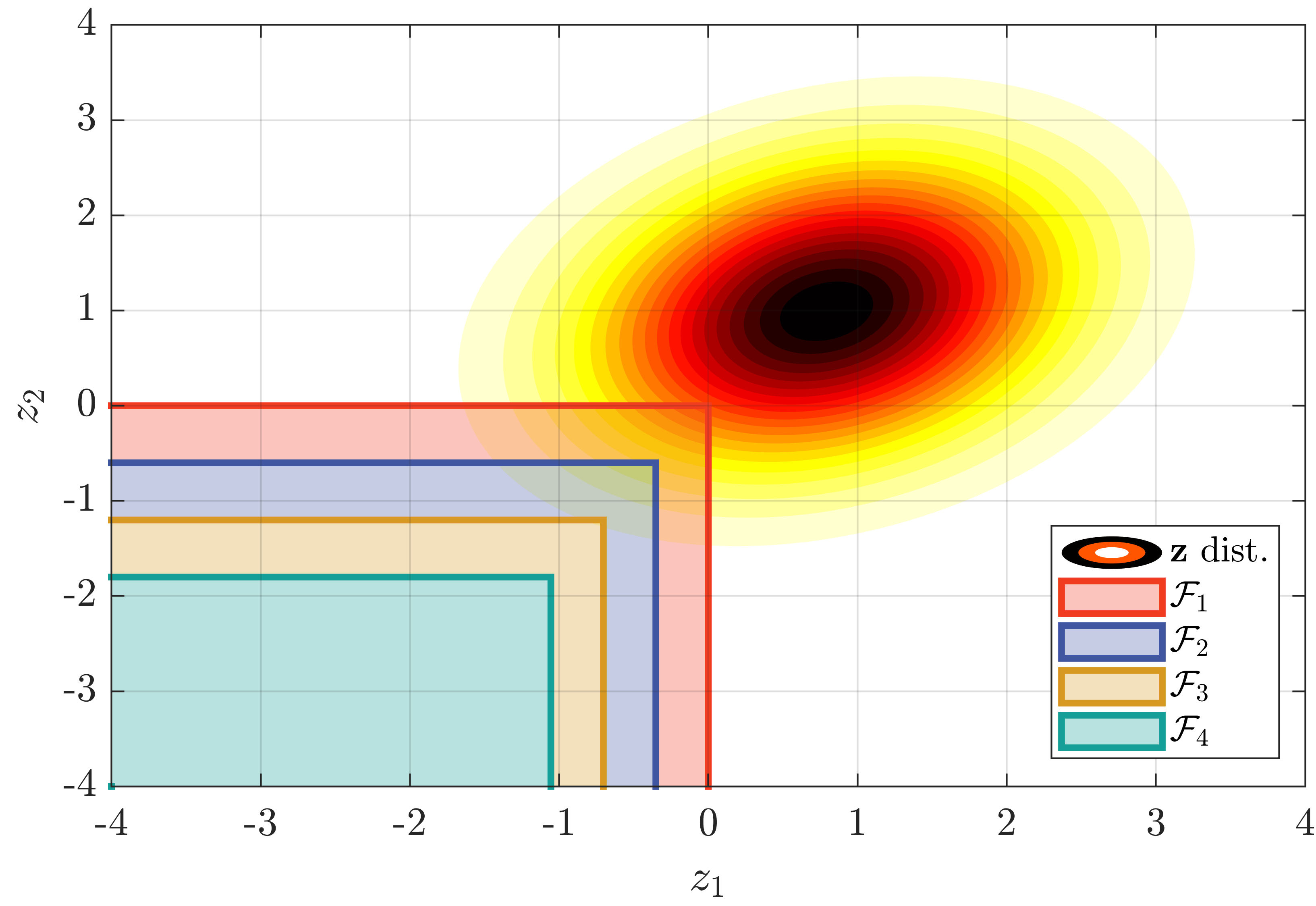}
        \caption{Overlaid color maps and intermediate failure domains obtained by linear transformation.}
        \label{Fig5}
    \end{figure}

    Let $F_k$ be the network failure domain under an earthquake with the $k^\text{th}$ moment magnitude $M_w^k$, where $M_w^1>M_w^2>\dots>M_w^m$. Owing to the constant $\mathbf{R}_\mathbf{zz}$, all distributions of $\vect{\mathbf{z}}\left(M_w^k\right)$ can be matched to that of $\vect{\mathbf{z}}\left(M_w^1\right)$ by a linear transformation. Note that, unlike the conventional subset simulation where a relaxation parameter can be explicitly introduced to yield $\left\{G\left(\vect{\mathbf{z}}\left(M_w^k\right)\right)\leq g_k\right\}$, the network LSF is nonlinear and a \say{$g_k$} term cannot be factorized out of $G(\cdot)$. Therefore, we define $F_k$ as
    \begin{equation}\label{Eq25}
        \begin{aligned}
            F_k=&\left\{G\left(\vect{\mathbf{z}}\left(M_w^k\right)\right)\leq0\right\}\\=&\left\{G\left(\vect{\mathbf{z}}\left(M_w^1\right)+\vect{\mathbf{z}}\left(M_w^k\right)-\vect{\mathbf{z}}\left(M_w^1\right)\right)\leq0\right\}\\=&\left\{G\left(\vect{\mathbf{z}}\left(M_w^1\right)+(k-1)\Delta\vect{\mathbf{z}}\right)\leq0\right\},
        \end{aligned}
    \end{equation} 
    where $k=1,2,\ldots,m$, and $G(\cdot)$ denotes a network LSF, such as $G_\text{OD}$, $G_k$, and $G_{k/N}$. Here, we enforce a constant magnitude decrement, i.e., $\Delta M_w=M_w^{k+1}-M_w^k$ is set to a negative constant. According to Eq.~\eqref{Eq3}, $\vect{\mathbf{z}}(M_w)$ is a linear function of $M_w$; therefore, $\Delta\vect{\mathbf{z}}=\vect{\mathbf{z}}\left(M_w^k\right)-\vect{\mathbf{z}}\left(M_w^{k-1}\right)$ is also a constant vector. The last line of Eq.~\eqref{Eq25} presents an interpretable form of the intermediate failure domains for a \say{specialized} SS for the network fragility, an extension of the conventional SS with a relaxation parameter introduced into the LSF. It is seen from Eq.~\eqref{Eq25} that the relaxation parameter is $k\geq1$, and the effect of applying $k>1$ is to increase the safety margins for all network nodes. It is worth mentioning that Xian and Wang \cite{xian2024relaxation} investigated a more general variation of SS, yielding a family of sequential sampling methods that do not rely on nested intermediate failure domains. Fig.~\ref{Fig5} illustrates the intermediate failure domains transformed from Fig.~\ref{Fig4}.

    \subsubsection{Generation of network fragility curves}\label{sec332}
    \noindent Using Eq.~\eqref{Eq25}, a single implementation of the specialized SS can yield the network failure probabilities at multiple values of $M_w$, i.e., the fragility curve. This practice requires significantly fewer samples than are required by crude MCS or by repeated SS runs for each $M_w$. If a conditional probability $\Prob{\mathcal{F}_i|\mathcal{F}_{i-1}}$ for a pre-specified magnitude decrement is too small, similar to the conventional SS, we can adaptively reduce the decrement of $M_w$ so that the conditional probability becomes large, i.e., building an adaptive mesh refinement for the fragility curve.

    \section{Numerical examples}\label{sec4}
    \noindent Three numerical examples are considered to demonstrate the efficiency and accuracy of the proposed network LSFs and the method for network fragility curve evaluation: (1) two-terminal reliability on the two-component parallel system, (2) $k$-terminal reliability on the San Jose highway bridge network \cite{guo2017seismic,nabian2018deep,lee2021multi}, and (3) $k$-out-of-$N$ reliability on the San Diego highway bridge network \cite{lee2021multi}. It is important to note that the numerical models for network reliability are simplifications of reality and are susceptible to various epistemic uncertainties, such as the variability in fragility functions and earthquake ground motions. A refined study aimed at reducing epistemic uncertainty would involve collecting infrastructure and seismicity data to update the network reliability models and their predictions. In each of these examples, the seismic capacity parameters for the components or bridges are fixed at 0.98 for the median ${\bar{C}}_i$, and 0.69 for the log-standard deviation $\zeta_i$. However, one can vary these parameters for each structure, accounting for their lifespan and target damage state, such as slight, moderate, extensive, complete damage.

    For the two proposed network LSFs, $G_\text{OD}^\text{RP}$ and $G_\text{OD}^\text{SP}$, the parameters for HMC-SS are set to $n=1,000$, $p_0=0.1$, $\smash{t_f}=\pi/4$, and $\alpha=0$ (for details on the last two parameters, see \cite{wang2019hamiltonian}). Their performances are compared with other simulation-based approaches (e.g., sequential space conversion (SESC) \cite{rashki2021sesc}, adaptive effort subset simulation (aE-SuS) \cite{chan2022adaptive}, MCS) using Eq.~\eqref{Eq12}, in terms of sample efficiency, accuracy, and computational time. The parameters for SESC are set to $A = [2,1.95,\ldots,1]$ and $N_f=100$ (for details, see \cite{rashki2021sesc}) by default, with $N_f$ increased until the SESC works without accuracy loss when the target probability is low. For aE-SuS, the parameters are set to $tol=0.5$, $p_0=0.1$ (for details, see \cite{chan2022adaptive}). The MCS solution is obtained through crude MCS with a target c.o.v., $\delta_{\smash{\hat{P}_f}}=0.01$. All computations in this section are performed using MATLAB\textregistered{} on an 8-core MacBook Air (2022) with 8 GB of RAM.
    
    \subsection{Example 1: a two-component parallel system}\label{sec41}
\begin{figure}[H]
        \centering
        \includegraphics[width=\textwidth]{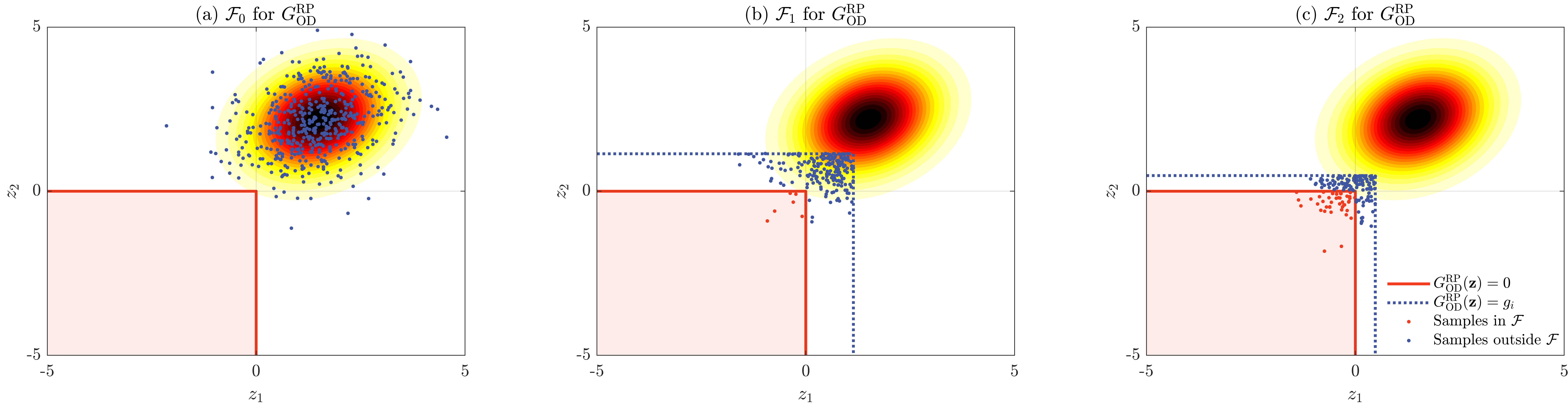}
        \caption{Samples obtained from the (a) entire domain; (b) first subset; and (c) second subset using $G_\text{OD}^\text{RP}$.}
        \label{Fig6}
    \end{figure}

\begin{figure}[H]
        \centering
        \includegraphics[width=\textwidth]{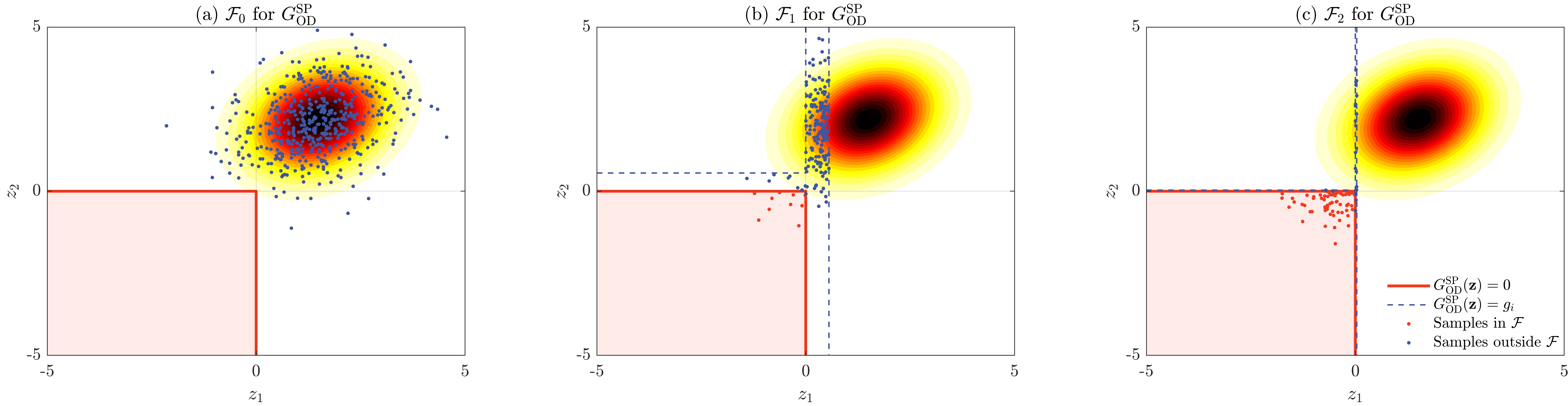}
        \caption{Samples obtained from the (a) entire domain; (b) first subset; and (c) second subset using $G_\text{OD}^\text{SP}$.}
        \label{Fig7}
    \end{figure}

    \noindent Consider the two-terminal reliability of the two-component parallel system mentioned above (see \ref{A1}). For an earthquake with $M_w=5.0$, Figs.~\ref{Fig6} and \ref{Fig7} represent the 500 HMC samples and the adaptively identified intermediate failure domains using $G_\text{OD}^\text{RP}$ and $G_\text{OD}^\text{SP}$, respectively. The contour plot of the joint PDF $f_{\vect{\mathbf{Z}}}(\vect{\mathbf{z}})$ and the system failure domain $\mathcal{F}$ are also shown. In contrast to the square intermediate domains of $G_\text{OD}^\text{RP}$, those of $G_\text{OD}^\text{SP}$ resemble the plots in Fig.~\ref{Fig3}(b).

    To evaluate the bias and variation in the results obtained with the method, HMC-SS is executed 500 times using $G_\text{OD}^\text{RP}$ and $G_\text{OD}^\text{SP}$. Table~\ref{tab:1} shows the results compared with the $P_f$ estimates from other simulation-based methods. The \say{exact} $P_f$ of the two-component system is estimated using a two-fold numerical integration, which is infeasible for large-scale networks. While the estimates of HMC-SS are asymptotically unbiased, the solutions of HMC-SS achieve high accuracy when compared with the exact $P_f$. This is also valid for SESC and aE-SuS which utilize SS, although many more samples are required for the methods at low $M_w$. Table~\ref{tab:1} also summarizes the c.o.v. of the estimates, $\delta_{\smash{\hat{P}_f}}$, the average number of network LSF evaluations, $N_G$, and the computation time per run, $t$.

    \begin{table}[H]
        \centering
        \caption{Two-terminal reliability analysis results for the two-component parallel system.}
        \label{tab:1}
        \begin{tabular}{l c c c c c c c c}
            \toprule
            \multirow{2}{*}{Method} & \multicolumn{4}{c}{$M_w=4.0$} & \multicolumn{4}{c}{$M_w=3.0$} \\
            \cmidrule(lr){2-5} \cmidrule(lr){6-9}
            & $\smash{\hat{P}_f}$ & $\delta_{\smash{\hat{P}_f}}$ & $N_G$ & $t$(s) & $\smash{\hat{P}_f}$ & $\delta_{\smash{\hat{P}_f}}$ & $N_G$ & $t$(s) \\
            \midrule
            HMC-SS ($G_\mathrm{OD}^\mathrm{RP}$) & $1.00 \times 10^{-3}$ & 0.255 & 3,290 & 0.0220 & $3.46 \times 10^{-4}$ & 0.286 & 3,700 & 0.0253 \\
            HMC-SS ($G_\mathrm{OD}^\mathrm{SP}$) & $1.00 \times 10^{-3}$ & 0.490 & 3,302 & 0.0073 & $3.42 \times 10^{-4}$ & 0.699 & 4,103 & 0.0092 \\
            SESC & $1.04 \times 10^{-3}$ & 0.532 & 10,119 & 0.0266 & $3.56 \times 10^{-4}$ & 0.541 & 23,018 & 0.110 \\
            aE-SuS & $0.99 \times 10^{-3}$ & 0.138 & 55,108 & 0.240 & $3.43 \times 10^{-4}$ & 0.125 & 168,626 & 0.752 \\
            Crude MCS & $1.00 \times 10^{-3}$ & 0.010 & 1.00E7 & 37.5 & $3.41 \times 10^{-4}$ & 0.010 & 2.93E7 & 112 \\
            Exact $P_f$ & $0.98 \times 10^{-3}$ & & & & $3.40 \times 10^{-4}$ & & & \\
            \bottomrule
        \end{tabular}%
    \end{table}

    We introduce $eff=\delta_{\smash{\hat{P}_f}}\times\sqrt{N_G}$ to measure the efficiency of the sampling methods \cite{au2001estimation}. A lower $eff$ indicates higher efficiency. Furthermore, the relative efficiency ($\text{relEff}$) with respect to the crude MCS allows us to further measure the efficiency gain \cite{lecuyer1994efficiency,chan2023bayesian}, which is defined as
    \begin{equation}\label{Eq26}
        \text{relEff}\left(\smash{\hat{P}_f}\right)=\frac{P_f(1-P_f)}{\text{MSE}\left(\smash{\hat{P}_f}\right)\times N_G},
    \end{equation}
    where $\text{MSE}(\smash{\hat{P}_f})$ is the mean square error of the estimates $\smash{\hat{P}_f}$; and $P_f$ denotes the exact value, which is replaced by the MCS estimate if the analytic solution is infeasible. The relative efficiency of the crude MCS converges to 1, with a higher $\text{relEff}$ representing superior performance in terms of accuracy and efficiency.

\begin{figure}[htbp]
        \centering
        \includegraphics[width=\textwidth]{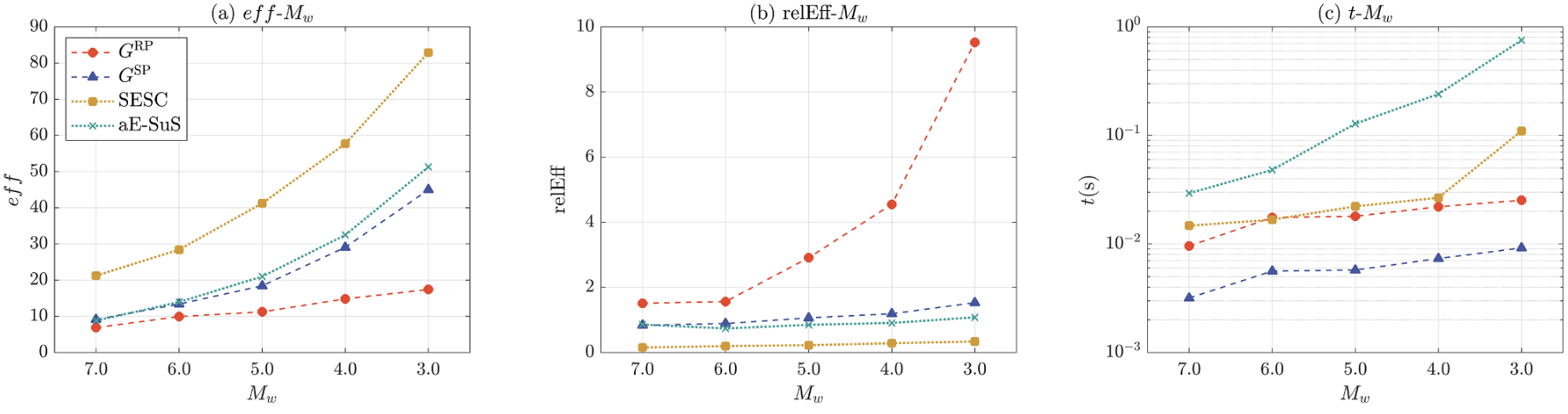}
        \caption{(a) $eff$; (b) $\text{relEff}$; and (c) $t$ as functions of $M_w$ for the two-component parallel system.}
        \label{Fig8}
    \end{figure}

    Fig.~\ref{Fig8}(a) to (c) compare the efficiency of the proposed network LSFs with other simulation-based methods in terms of $eff$, $\text{relEff}$, and the computation time while varying $M_w$. For all methods, as $M_w$ decreases, the estimated network failure probabilities decrease, and $eff$ and $t$ increase. This is an intrinsic property of simulation-based methods, but the rate of increase varies among methods. It is notable that the proposed LSF-based HMC-SS outperforms the other binary LSF-based methods across all evaluated metrics. Fig.~\ref{Fig8}(a) shows that the HMC-SS employing the proposed LSFs has better efficiency. In Fig.~\ref{Fig8}(b), the $\text{relEff}$ of $G_\text{OD}^\text{RP}$-based HMC-SS increases sharply for low $M_w$, indicating that the efficiency gain of the proposed method is significant when the target probability is low. In contrast, the SESC and aE-SuS using the binary LSF have lower efficiency in estimating small probabilities. For binary LSFs, aE-SuS is similar to the crude MCS before identifying samples in the target failure domain; thus, its efficiency gain over MCS is marginal. It is also found that the $G_\text{OD}^\text{RP}$-based HMC-SS is more accurate (low $\delta_{\smash{\hat{P}_f}}$), whereas the $G_\text{OD}^\text{SP}$-based one is more efficient for all values of $M_w$. The speedup associated with $G_\text{OD}^\text{SP}$ comes from the efficient BFS-based shortest path search, which is considerably faster than Dijkstra's algorithm used in $G_\text{OD}^\text{RP}$. However, despite the utilization of Dijkstra's algorithm, $G_\text{OD}^\text{RP}$-based HMC-SS outperforms binary LSF-based methods, where BFS is used.

\begin{figure}[htbp]
        \centering
         \includegraphics[width=.75\textwidth]{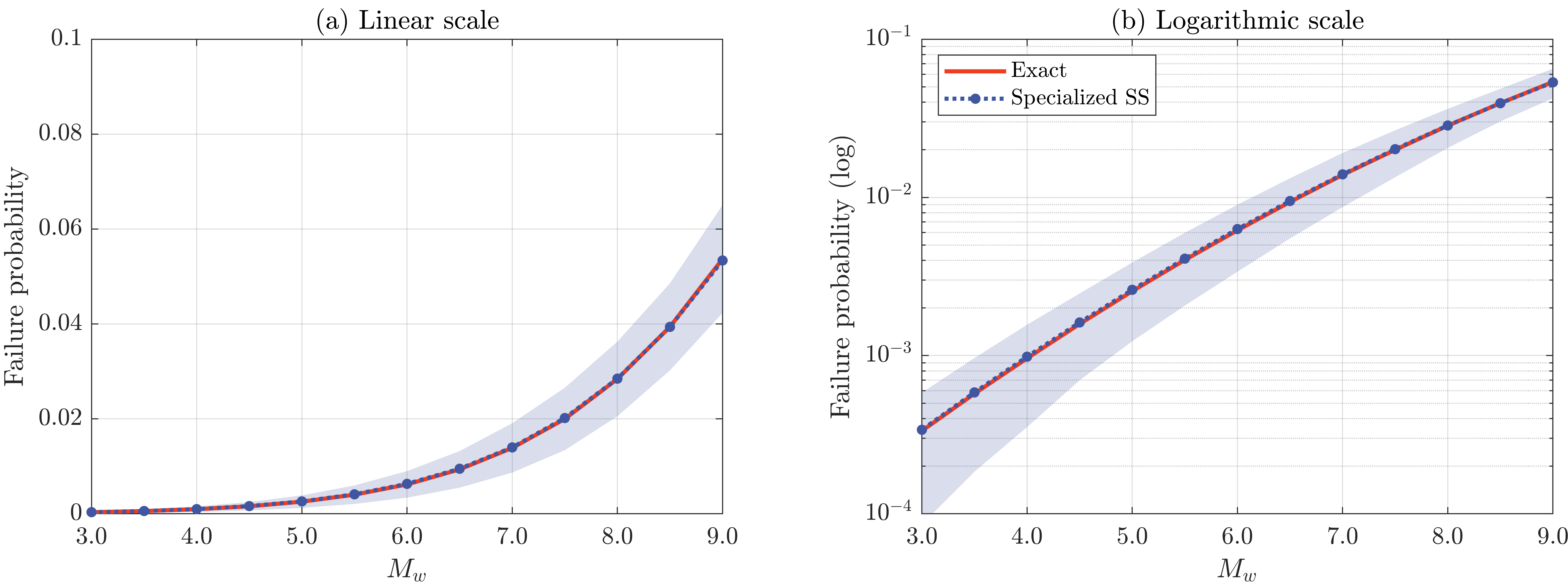}
        \caption{Seismic system fragility curves obtained by a single implementation of specialized HMC-SS.}
        \label{Fig9}
    \end{figure}

    Next, we evaluate the seismic fragility curve using the framework proposed in Section \ref{sec33}. In a single implementation of the specialized HMC-SS, $G_\text{OD}^\text{RP}$ is adopted for its high accuracy, even though using it requires more computation time than using $G_\text{OD}^\text{SP}$. The range of $M_w$ is set to $3.0 \leq M_w \leq 9.0$ with $\Delta M_w = 0.5$, and the specialized HMC-SS is repeated 250 times to produce an estimate of the confidence interval. Fig.~\ref{Fig9} shows the generated seismic fragility curve (blue dashed line) along with the 95\% confidence interval (blue shaded area) compared to the exact values (red solid line). The specialized HMC-SS estimates the fragility curve accurately using 12,700 LSF evaluations, which is only 37.47\% of those required in a repeated simulation of HMC-SS for each $M_w$. Appendix~\ref{B1} provides the detailed results of the \say{one-shot} HMC-SS compared to those from separate runs.

    \subsection{Example 2: San Jose highway bridge network}\label{sec42}
\begin{figure}[tbp]
        \centering
        \includegraphics[width=.33\textwidth]{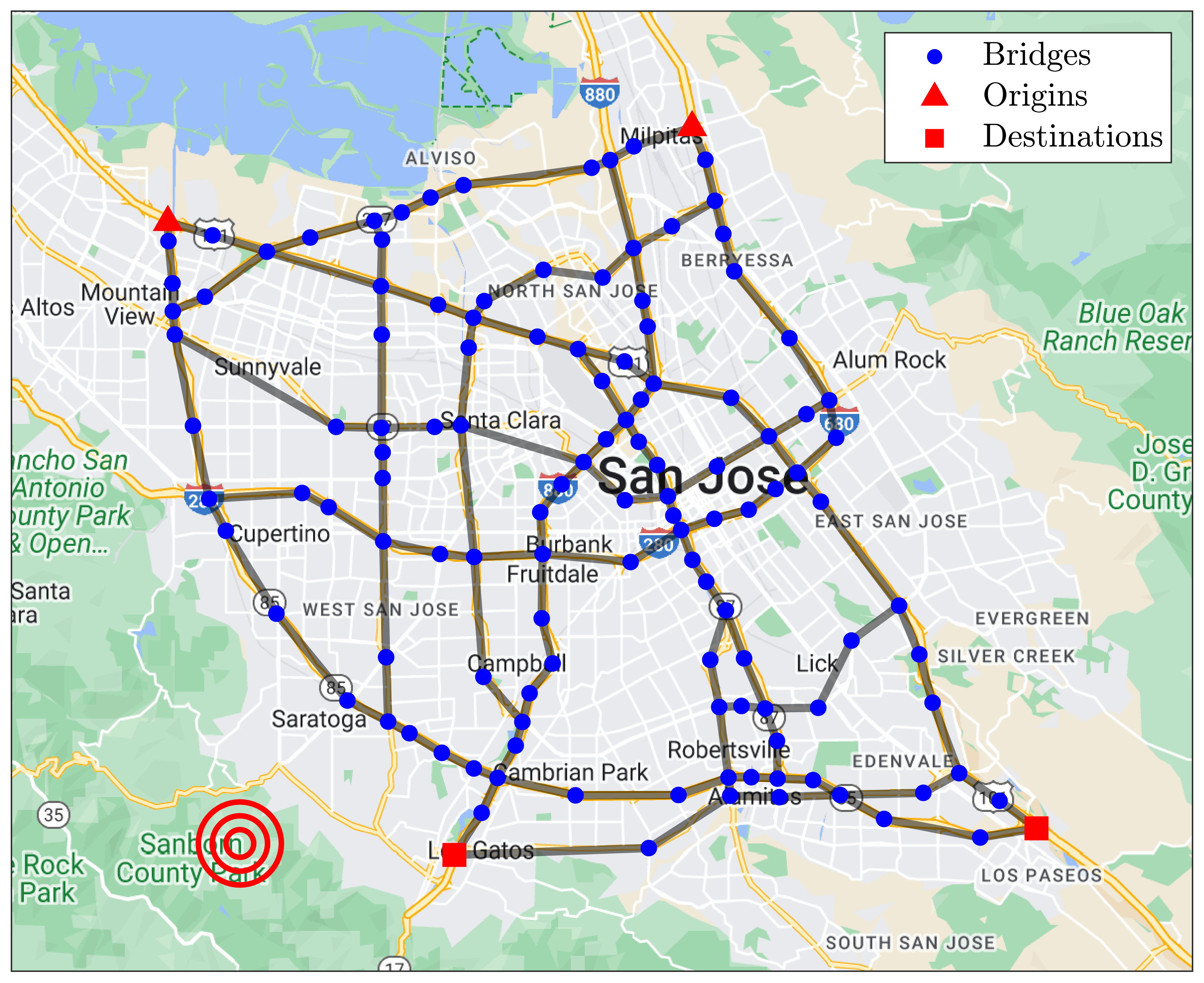}
        \caption{Highway bridge network in San Jose, CA \cite{lee2021multi}.}
        \label{Fig10}
    \end{figure}

\begin{figure}[htbp]
        \centering
        \includegraphics[width=\textwidth]{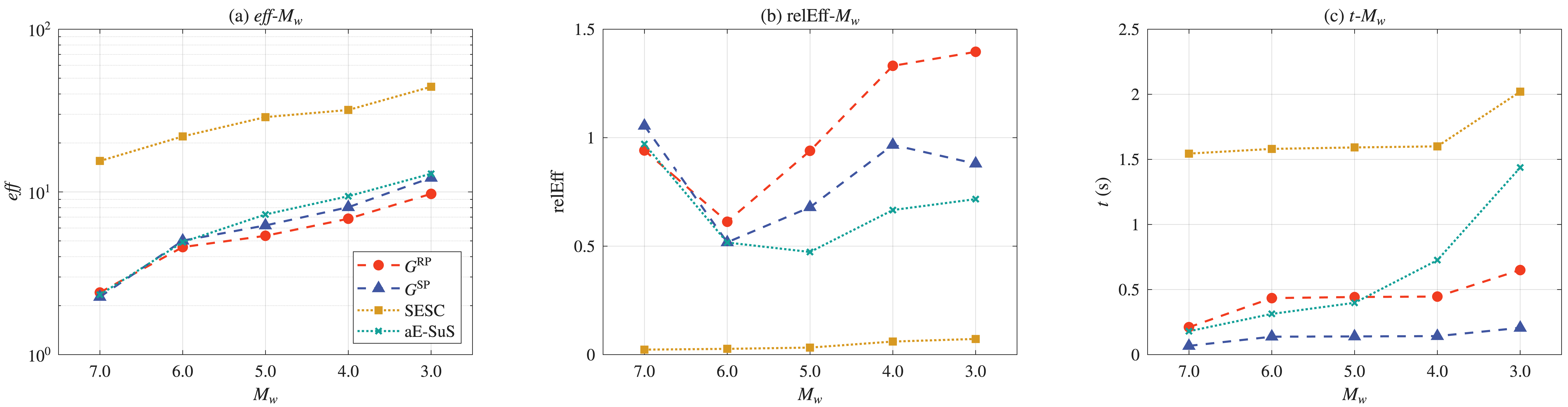}
        \caption{(a) $eff$-$M_w$; (b) $\text{relEff}$-$M_w$; and (c) $t$-$M_w$ curves for the San Jose highway bridge network.}
        \label{Fig11}
    \end{figure}

    \noindent Fig.~\ref{Fig10} shows the highway bridge network in San Jose, California \cite{lee2021multi} (modified from \cite{guo2017seismic,nabian2018deep}) with two origins and two destinations. To analyze the $k$-terminal reliability (in this example, $k = \left| \vect{V}_{O} \right| + \left| \vect{V}_{D} \right| = 4$), HMC-SS using $G_\text{OD}^\text{RP}$ and $G_\text{OD}^\text{SP}$, along with other simulation-based approaches, is conducted 500 times, and Table~\ref{tab:2} summarizes the performance. The results confirm the accuracy of the HMC-SS using the proposed network LSFs. Fig.~\ref{Fig11} compares $eff$, $\text{relEff}$, and $t$ obtained from the two proposed LSFs with other methods. Similar to Example 1, the trade-off between accuracy and efficiency seems inevitable in HMC-SS with the proposed LSFs. In this example, the efficiency of the proposed method is not as pronounced compared to the crude MCS because $\smash{\hat{P}_f}$ is relatively higher than those in Example 1 at the same earthquake magnitude. However, when $M_w\leq4.0$, the RP-based HMC-SS outperforms the MCS and other methods in terms of efficiency. On the other hand, in terms of computation time, the SP-based approach outperforms the others regardless of the earthquake magnitude.

    \begin{table}[tbp]
    \centering
    \caption{Four-terminal reliability analysis results for the San Jose highway bridge network.}
    \label{tab:2}
    \resizebox{\linewidth}{!}{%
    \begin{tabular}{l c c c c c c c c}
        \toprule
        \multirow{2}{*}{Method} & \multicolumn{4}{c}{$M_w=4.0$} & \multicolumn{4}{c}{$M_w=3.0$} \\
        \cmidrule(lr){2-5} \cmidrule(lr){6-9}
        & $\smash{\hat{P}_f}$ & $\delta_{\smash{\hat{P}_f}}$ & $N_G$ & $t$(s) & $\smash{\hat{P}_f}$ & $\delta_{\smash{\hat{P}_f}}$ & $N_G$ & $t$(s) \\
        \midrule
        HMC-SS ($G_\mathrm{OD}^\mathrm{RP}$) & $1.59 \times 10^{-2}$ & 0.157 & 1,904 & 0.446 & $7.53 \times 10^{-3}$ & 0.185 & 2,760 & 0.650 \\
        HMC-SS ($G_\mathrm{OD}^\mathrm{SP}$) & $1.58 \times 10^{-2}$ & 0.184 & 1,916 & 0.142 & $7.55 \times 10^{-3}$ & 0.234 & 2,730 & 0.206 \\
        SESC & $1.60 \times 10^{-2}$ & 0.371 & 7,396 & 1.60 & $7.24 \times 10^{-3}$ & 0.503 & 7,741 & 2.02 \\
        aE-SuS & $1.62 \times 10^{-2}$ & 0.139 & 4,568 & 0.726 & $7.75 \times 10^{-3}$ & 0.135 & 9,226 & 1.44 \\
        Crude MCS & $1.56 \times 10^{-2}$ & 0.010 & 6.13E5 & 38.0 & $7.56 \times 10^{-3}$ & 0.010 & 1.32E6 & 83.8 \\
        \bottomrule
    \end{tabular}%
    }
    \end{table}
    
\begin{figure}[htbp]
        \centering
        \includegraphics[width=.75\textwidth]{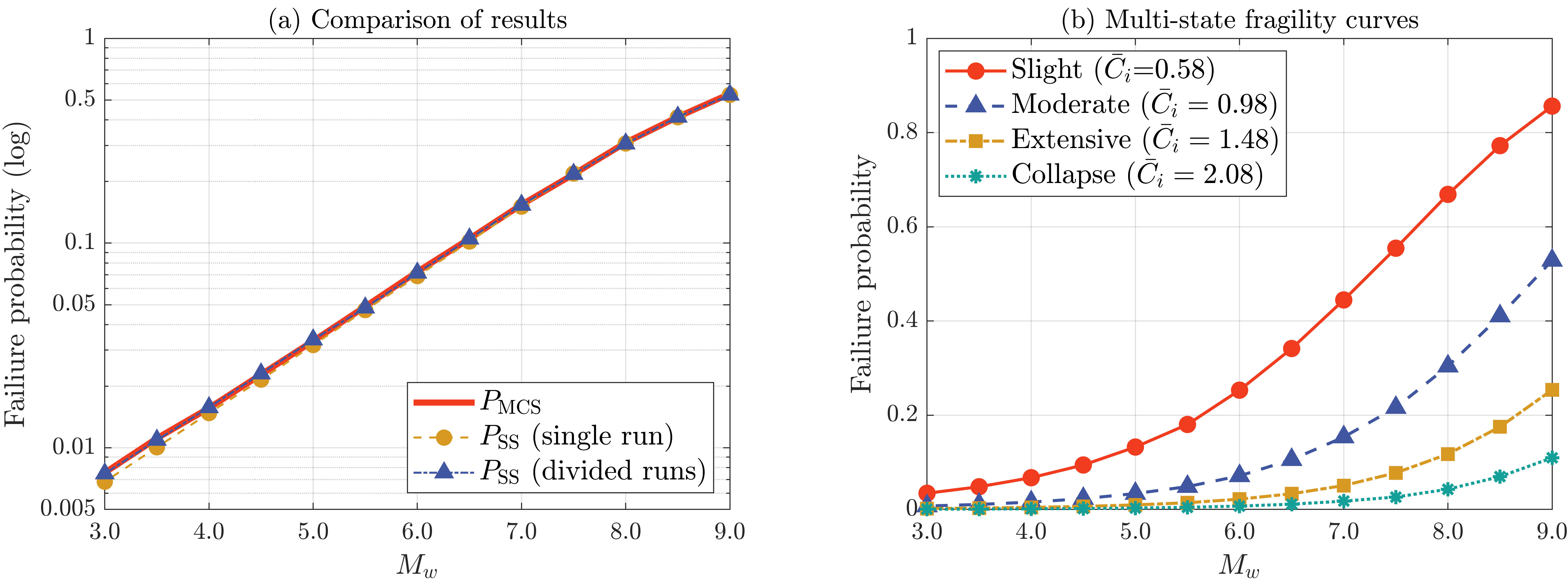}
        \caption{(a) Comparison of generated seismic fragility curves of the San Jose highway bridge network; and (b) multi-state network fragility curves obtained by specialized HMC-SS.}
        \label{Fig12}
    \end{figure}

    By substituting Eq.~\eqref{Eq23} into Eq.~\eqref{Eq25}, the network fragility curve can be evaluated in terms of four-terminal reliability as a function of magnitude. Accumulated biases/errors of sequential conditioning are particularly conspicuous when assessing $k$-terminal or $k$-out-of-$N$ reliability in large-scale networks because of the complexity of failure domains. To minimize the error of specialized subset simulations due to accumulated bias, we divide the target range of $M_w$ into multiple intervals, where independent specialized SSs are performed for each interval. Each run adaptively selects the intermediate failure domain using the $p_0( = 0.1)$-quantile until the samples reach the first failure domain of the interval. This approach can be understood as using a coarse adaptive mesh controlled by $p_0$ to reach the target interval and a uniform fine mesh controlled by $\Delta M_w$ to generate a smooth fragility curve within the target interval. Compared with a \say{one-shot} run with a constant $\Delta M_w$ to produce the entire fragility curve, this \say{divide-and-conquer} trick requires fewer steps to reach each interval; thus, the bias is smaller, at the cost of using more sample points. The range of $3.0 \leq M_w \leq 9.0$ is divided into three intervals: $3.0 \leq M_w \leq 5.0$, $5.0 < M_w \leq 7.0$, and $7.0 < M_w \leq 9.0$, with $\Delta M_w = 0.5$. In this example, HMC-SS is executed 250 times. Fig.~\ref{Fig12}(a) compares the results of the crude MCS, the single implementation of the specialized HMC-SS, and three divided implementations of the specialized HMC-SS. Note that the divided run provides more accurate results consistent with the crude MCS result, while the single run underestimates the failure probabilities when $M_w \leq 5.0$ due to accumulation of biases/errors. To achieve better accuracy, the divided runs require only 9.32\% more samples than the single run. The detailed results of the specialized HMC-SS are presented in Appendix~\ref{B2}. In addition, the proposed framework can generate multi-state fragility curves by varying the seismic capacity parameters, as shown in Fig.~\ref{Fig12}(b). All bridges in the network are assumed to have the following median seismic capacities ${\bar{C}}_i$ for each damage state: 0.58 for slight damage, 0.98 for moderate damage, 1.48 for extensive damage, and 2.08 for complete damage. For each damage state, the log-standard deviation $\zeta_i$ is set to 0.69.

    \subsection{Example 3: San Diego highway bridge network}\label{sec43}
    \begin{figure}[htbp]
    \centering
    \includegraphics[width=.33\textwidth]{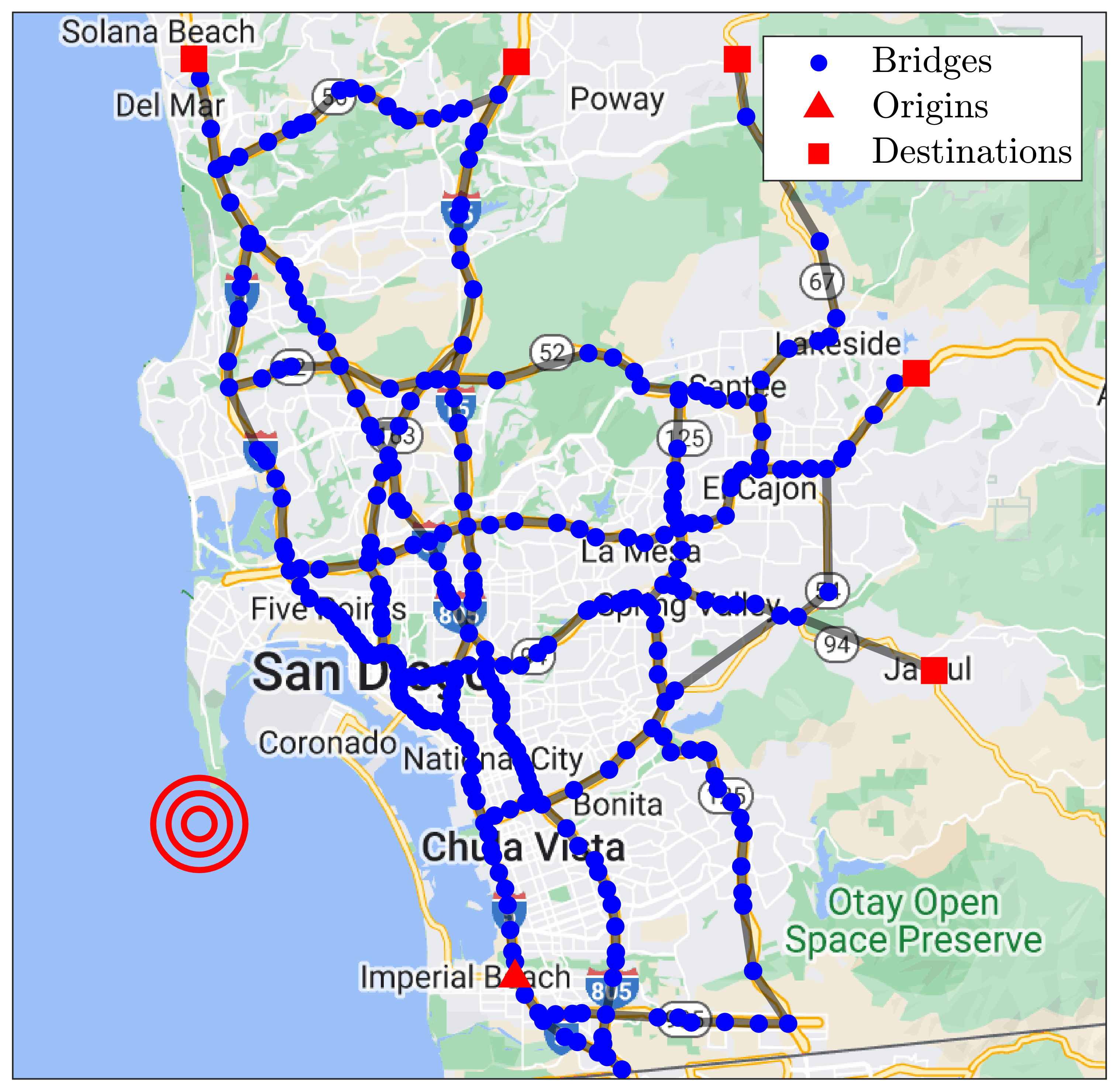}
    \caption{Highway bridge network in San Diego, CA \cite{lee2021multi}.}
    \label{Fig13}
    \end{figure}

\begin{figure}[H]
    \centering
    \includegraphics[width=\textwidth]{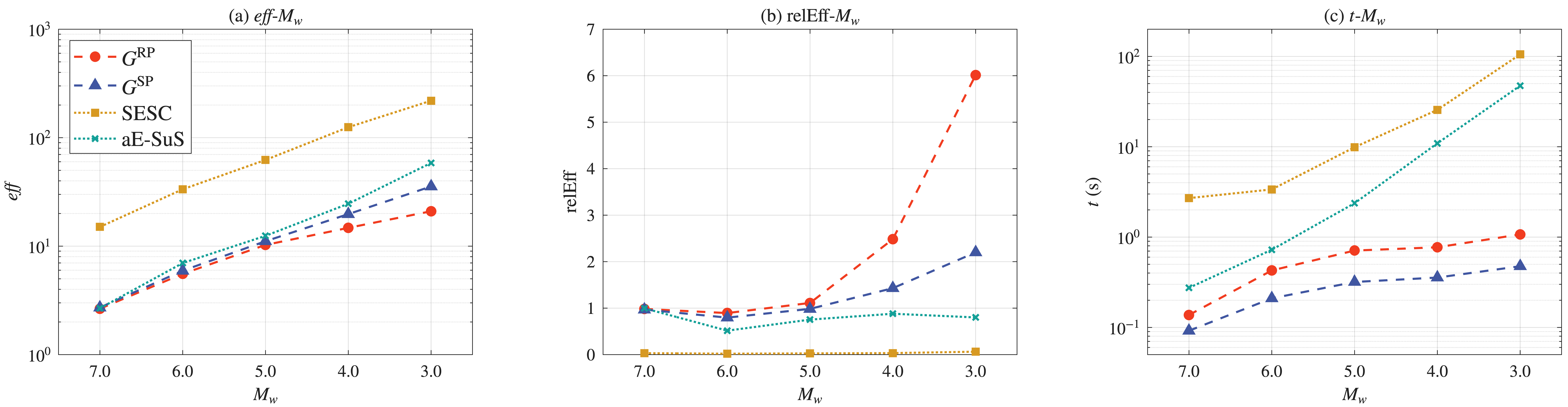}
    \caption{(a) $eff$-$M_w$; (b) $\text{relEff}$-$M_w$; and (c) $t$-$M_w$ curves for the San Diego highway bridge network.}
    \label{Fig14}
    \end{figure}

    \noindent Fig.~\ref{Fig13} shows the highway bridge network in San Diego, California \cite{lee2021multi} with one origin and five destinations, connecting the southwest coast of San Diego to five nearby cities. Eq.~\eqref{Eq24} is used to evaluate the probability of at least $k \leq 5$ destinations being accessible from the origin (i.e., the $k$-out-of-5 reliability). Table \ref{tab:3} provides a comparison of the 3-out-of-5 reliability results estimated from HMC-SS using $G_\text{OD}^\text{RP}$ and $G_\text{OD}^\text{SP}$, and from other simulation-based methods using $G_\text{OD}^\text{Bi}$. Additionally, Fig.~\ref{Fig14} visualizes the $eff$-$M_w$, $\text{relEff}$-$M_w$, and $t$-$M_w$ curves. Similar to the previous examples, higher sampling efficiency and lower computation time of the proposed LSFs are confirmed compared to other sampling methods, and the trade-off between accuracy and efficiency is observed.

    \begin{table}[H]
        \centering
        \caption{3-out-of-5 reliability analysis results for the San Diego highway bridge network.}
        \label{tab:3}
        \resizebox{\linewidth}{!}{%
        \begin{tabular}{l c c c c c c c c}
            \toprule
            \multirow{2}{*}{Method} & \multicolumn{4}{c}{$M_w=4.0$} & \multicolumn{4}{c}{$M_w=3.0$} \\
            \cmidrule(lr){2-5} \cmidrule(lr){6-9}
            & $\smash{\hat{P}_f}$ & $\delta_{\smash{\hat{P}_f}}$ & $N_G$ & $t$(s) & $\smash{\hat{P}_f}$ & $\delta_{\smash{\hat{P}_f}}$ & $N_G$ & $t$(s) \\
            \midrule
            HMC-SS ($G_\mathrm{OD}^\mathrm{RP}$) & $1.75\times10^{-3}$ & 0.277 & 2,841 & 0.771 & $3.69\times10^{-4}$ & 0.345 & 3,702 & 1.07 \\
            HMC-SS ($G_\mathrm{OD}^\mathrm{SP}$) & $1.73\times10^{-3}$ & 0.366 & 2,904 & 0.357 & $3.61\times10^{-4}$ & 0.580 & 3,758 & 0.476 \\
            SESC & $1.81\times10^{-3}$ & 0.524 & 56,905 & 25.6 & $3.43\times10^{-4}$ & 0.528 & 172,566 & 105 \\
            aE-SuS & $1.87\times10^{-3}$ & 0.330 & 8,846 & 2.06 & $3.63\times10^{-4}$ & 0.350 & 44,466 & 11.7 \\
            Crude MCS & $1.69\times10^{-3}$ & 0.010 & 6.09E6 & 200 & $3.62\times10^{-4}$ & 0.010 & 2.92E7 & 1,008 \\
            \bottomrule
        \end{tabular}%
        }
    \end{table}

\begin{figure}[htbp]
    \centering
    \includegraphics[width=0.75\textwidth]{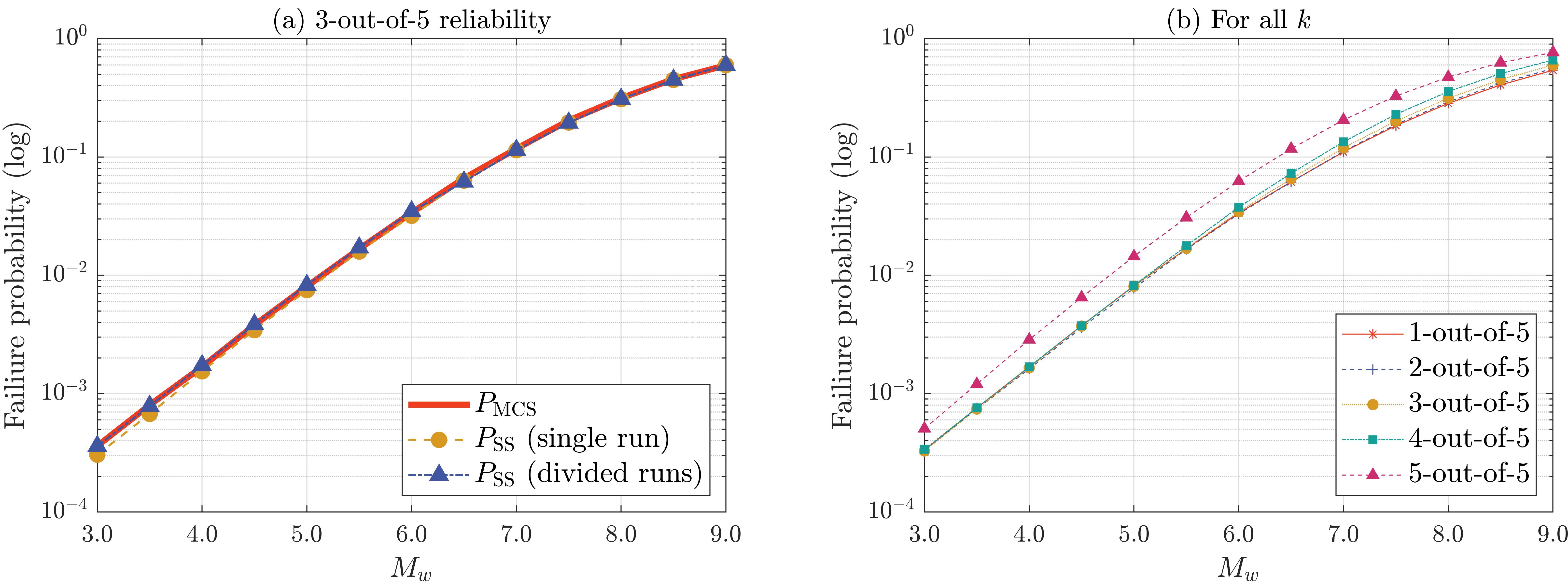}
    \caption{Seismic fragility curves of the San Diego highway bridge network for (a) $k=3$; and (b) $1\leq k \leq5$.}
    \label{Fig15}
    \end{figure}

    For fragility analysis, the range of $3.0\leq M_w\leq9.0$ is divided into two intervals, i.e., $3.0 \leq M_w \leq 6.0$, and $6.0 < M_w \leq 9.0$, with $\Delta M_w=0.5$, and HMC-SS is repeated 250 times. Fig.~\ref{Fig15}(a) shows the results of the 3-out-of-5 reliability evaluated by HMC-SS with and without division, in comparison to those by the crude MCS. As in Example 2, one can observe that the divided implementations of the HMC-SS can avoid underestimating the failure probability at low $M_w$, unlike the single implementation. The detailed results of the 3-out-of-5 reliability evaluated by the specialized HMC-SS are provided in Appendix~\ref{B3}. By repeating this assessment process for $1 \leq k \leq 5$, we obtain the fragility curves shown in Fig.~\ref{Fig15}(b), where we increase the value of $n$ to 10,000 to achieve more stable convergence.

    \section{Conclusions}\label{sec5}
    \noindent Two informative limit-state functions (LSFs) were proposed to reformulate the traditional binary LSF for two-terminal reliability analysis of lifeline networks, thereby making the application of subset simulation (SS) possible. The first LSF quantifies the vulnerability of the most reliable path between the origin and destination nodes, whereas the other utilizes the shortest path. Both LSF reformulations represent the same failure domain as the original binary LSF, while their function values in the safe domain differ. Moreover, a specialized SS can be developed to generate network-level fragility curves by connecting intermediate failure events to the earthquake magnitude. As a result, the specialized SS can generate the network fragility curve in a single run. Furthermore, the sampling framework was successfully extended to assess $k$-terminal or $k$-out-of-$N$ reliability.

    The accuracy and efficiency of the proposed framework were tested and demonstrated by three numerical examples using the Hamiltonian Monte Carlo-based SS. The results confirm the accuracy and efficiency of the proposed network LSFs, compared to other simulation-based methods directly applicable to binary LSFs. The first LSF generally has higher accuracy, while the second requires less computation time.

    As demonstrated by numerical examples, the proposed framework is highly scalable and can be applied to generating seismic fragility curves for various network reliability indices. Future research could develop informative LSFs tailored to each network reliability index instead of relying solely on the proposed functions that measure the vulnerability of the most reliable or shortest paths. Another possibility for future research would be to propose an alternative LSF applicable to other sampling methods. A remaining task is to use the proposed framework to identify more realistic network reliability, such as network flow capacity. Possible solutions to this problem could involve modeling components as multi-state or continuous, or considering the flow capacity of each origin-destination pair in the proposed $k$-out-of-$N$ reliability analysis. Finally, probabilistic inferences, like sensitivity analysis or importance measures for each component, can be conducted.

    \section*{CRediT authorship contribution statement}
    \noindent \textbf{Dongkyu Lee:} Writing -- original draft, Visualization, Validation, Software, Methodology, Conceptualization. \textbf{Ziqi Wang:} Writing -- review \& editing, Visualization, Validation, Supervision, Software, Methodology, Conceptualization. \textbf{Junho Song:} Writing -- review \& editing, Supervision, Resources, Project administration, Funding acquisition, Conceptualization.

    \section*{Declaration of competing interest}
    \noindent The authors declare that they have no known competing financial interests or personal relationships that could have appeared to influence the work reported in this paper.

    \section*{Acknowledgment}
    \noindent This work is supported by the Korea Agency for Infrastructure Technology Advancement (KAIA) grant funded by the Ministry of Land, Infrastructure and Transport (RS-2021-KA163162). The first author is also supported by the Basic Science Research Program through the National Research Foundation of Korea (NRF) funded by the Ministry of Education (RS-2024--00408907). The third author is also supported by the Institute of Construction and Environmental Engineering at Seoul National University. This support is gratefully acknowledged.
    \section*{Data availability}
    \noindent Data will be made available on request.
    
    \bibliography{RefList}

    \appendix
    \section{Details of the two-component parallel system}\label{A1}
    \noindent The distance $\Delta_{12}$ between the two components is given as 11.12 km, and the distances $R_1$ and $R_2$ from the epicenter to the two components are given as 3.46 km and 9.28 km, respectively. For all components, the seismic capacity parameters are assumed to be 0.98 for the median $\bar{C}_i$ and 0.69 for the log-standard deviation $\zeta_i$.

    According to Eqs.~\eqref{Eq12}--\eqref{Eq13}, the binary network LSF is given as $G_\text{parallel}^\text{Bi}=1-\mathbb{I}(z_1\leq0)\mathbb{I}(z_2\leq0)$, where the reliability index vector and the covariance matrix are calculated as $\vect{\beta}=[-0.176M_w+2.376,-0.300M_w+3.693]\tr$ and $\mathbf{R}_{\mathbf{zz}}=\begin{bmatrix}1&0.2435\\0.2435&1\end{bmatrix}$, respectively, from Eqs.~\eqref{Eq2}--\eqref{Eq4} and \eqref{Eq8}.

    \section{Detailed results of numerical examples in Section \ref{sec4}}\label{B}
    \setcounter{table}{0}
    \renewcommand{\thetable}{B\arabic{table}}
    \renewcommand{\theHtable}{B\arabic{table}}

    \begin{table}[!htbp]
        \centering
        \caption{Details of two-terminal reliability estimates for the two-component parallel system.}\label{B1}
    \begin{tabular}{c c c c c c c c}
        \toprule
        \multirow{2}{*}{$M_w$} & \multicolumn{3}{c}{Separate implementations} & \multicolumn{3}{c}{Single implementation} & \multirow{2}{*}{Exact $P_f$} \\ \cmidrule(lr){2-4} \cmidrule(lr){5-7}
        & $\smash{\hat{P}_{f,SS}}$ & $\delta_{\smash{\hat{P}_f}}$ & $N_{G}$ & $\smash{\hat{P}_{f,SS}}$ & $\delta_{\smash{\hat{P}_f}}$ & $N_{G}$ \\ \midrule
        9.0 & $5.38\times10^{-2}$ & 0.110 & 1,900 & $5.37\times10^{-2}$ & 0.108 & 1,900 & $5.34\times10^{-2}$ \\
        8.5 & $3.99\times10^{-2}$ & 0.120 & 1,900 & $3.94\times10^{-2}$ & 0.119 & 900 & $3.94\times10^{-2}$ \\
        8.0 & $2.84\times10^{-2}$ & 0.138 & 1,900 & $2.84\times10^{-2}$ & 0.141 & 900 & $2.85\times10^{-2}$ \\
        7.5 & $2.04\times10^{-2}$ & 0.145 & 1,900 & $2.00\times10^{-2}$ & 0.168 & 900 & $2.02\times10^{-2}$ \\
        7.0 & $1.41\times10^{-2}$ & 0.149 & 1,927 & $1.39\times10^{-2}$ & 0.191 & 900 & $1.40\times10^{-2}$ \\
        6.5 & $9.55\times10^{-3}$ & 0.179 & 2,465 & $9.36\times10^{-3}$ & 0.210 & 900 & $9.49\times10^{-3}$ \\
        6.0 & $6.34\times10^{-3}$ & 0.196 & 2,791 & $6.19\times10^{-3}$ & 0.231 & 900 & $6.31\times10^{-3}$ \\
        5.5 & $4.15\times10^{-3}$ & 0.187 & 2,800 & $4.02\times10^{-3}$ & 0.248 & 900 & $4.10\times10^{-3}$ \\
        5.0 & $2.62\times10^{-3}$ & 0.196 & 2,800 & $2.55\times10^{-3}$ & 0.264 & 900 & $2.60\times10^{-3}$ \\
        4.5 & $1.62\times10^{-3}$ & 0.241 & 2,840 & $1.58\times10^{-3}$ & 0.285 & 900 & $1.62\times10^{-3}$ \\
        4.0 & $1.00\times10^{-3}$ & 0.255 & 3,290 & $0.96\times10^{-3}$ & 0.323 & 900 & $0.98\times10^{-3}$ \\
        3.5 & $5.89\times10^{-4}$ & 0.269 & 3,682 & $5.75\times10^{-4}$ & 0.346 & 900 & $5.85\times10^{-4}$ \\
        3.0 & $3.46\times10^{-4}$ & 0.286 & 3,700 & $3.35\times10^{-4}$ & 0.379 & 900 & $3.40\times10^{-4}$ \\ \midrule
        Sum & & & 33,895 & & & 12,700 & \\ \bottomrule
        \end{tabular}
    \end{table}

    \begin{table}[H]
        \centering
        \caption{Details of four-terminal reliability estimates for the San Jose highway bridge network.}\label{B2}
    \begin{tabular}{c c c c c c c c}
        \toprule
        \multirow{2}{*}{$M_w$} & \multicolumn{3}{c}{Single implementation} & \multicolumn{3}{c}{Three divided implementations} & \multirow{2}{*}{MCS $\smash{\hat{P}_f}$} \\ \cmidrule(lr){2-4} \cmidrule(lr){5-7}
        & $\smash{\hat{P}_{f,SS}}$ & $\delta_{\smash{\hat{P}_f}}$ & $N_{G}$ & $\smash{\hat{P}_{f,SS}}$ & $\delta_{\smash{\hat{P}_f}}$ & $N_{G}$ \\ \midrule
        9.0 & $5.30\times10^{-1}$ & 0.029 & 1,000 & $5.30\times10^{-1}$ & 0.029 & 1,000 & $5.36\times10^{-1}$ \\
        8.5 & $4.12\times10^{-1}$ & 0.055 & 900 & $4.13\times10^{-1}$ & 0.040 & 900 & $4.13\times10^{-1}$ \\
        8.0 & $3.07\times10^{-1}$ & 0.069 & 900 & $3.06\times10^{-1}$ & 0.055 & 900 & $3.10\times10^{-1}$ \\
        7.5 & $2.19\times10^{-1}$ & 0.086 & 900 & $2.17\times10^{-1}$ & 0.083 & 900 & $2.17\times10^{-1}$ \\
        7.0 & $1.51\times10^{-1}$ & 0.118 & 900 & $1.54\times10^{-1}$ & 0.072 & 1,000 & $1.54\times10^{-1}$ \\
        6.5 & $1.02\times10^{-1}$ & 0.164 & 900 & $1.05\times10^{-1}$ & 0.121 & 900 & $1.05\times10^{-1}$ \\
        6.0 & $6.90\times10^{-2}$ & 0.203 & 900 & $7.15\times10^{-2}$ & 0.196 & 900 & $7.25\times10^{-2}$ \\
        5.5 & $4.71\times10^{-2}$ & 0.252 & 900 & $4.83\times10^{-2}$ & 0.279 & 900 & $4.90\times10^{-2}$ \\
        5.0 & $3.18\times10^{-2}$ & 0.319 & 900 & $3.38\times10^{-2}$ & 0.128 & 1,900 & $3.32\times10^{-2}$ \\
        4.5 & $2.15\times10^{-2}$ & 0.388 & 900 & $2.30\times10^{-2}$ & 0.180 & 900 & $2.28\times10^{-2}$ \\
        4.0 & $1.48\times10^{-2}$ & 0.452 & 900 & $1.58\times10^{-2}$ & 0.262 & 900 & $1.56\times10^{-2}$ \\
        3.5 & $1.01\times10^{-2}$ & 0.512 & 900 & $1.09\times10^{-2}$ & 0.339 & 900 & $1.11\times10^{-2}$ \\
        3.0 & $6.83\times10^{-3}$ & 0.569 & 900 & $7.48\times10^{-3}$ & 0.408 & 900 & $7.56\times10^{-3}$ \\ \midrule
        Sum & & & 11,800 & & & 12,900 & \\ \bottomrule
        \end{tabular}
    \end{table}  

    \begin{table}[H]
        \centering
        \caption{Details of 3-out-of-5 reliability estimates for the San Diego highway bridge network.}\label{B3}
    \begin{tabular}{c c c c c c c c}
        \toprule
        \multirow{2}{*}{$M_w$} & \multicolumn{3}{c}{Single implementation} & \multicolumn{3}{c}{Two divided implementations} & \multirow{2}{*}{MCS $\smash{\hat{P}_f}$} \\ \cmidrule(lr){2-4} \cmidrule(lr){5-7}
        & $\smash{\hat{P}_{f,SS}}$ & $\delta_{\smash{\hat{P}_f}}$ & $N_{G}$ & $\smash{\hat{P}_{f,SS}}$ & $\delta_{\smash{\hat{P}_f}}$ & $N_{G}$ \\ \midrule
        9.0 & $5.96\times10^{-1}$ & 0.026 & 1,000 & $5.96\times10^{-1}$ & 0.027 & 1,000 & $5.94\times10^{-1}$ \\
        8.5 & $4.48\times10^{-1}$ & 0.077 & 900 & $4.48\times10^{-1}$ & 0.045 & 900 & $4.55\times10^{-1}$ \\
        8.0 & $3.08\times10^{-1}$ & 0.094 & 900 & $3.09\times10^{-1}$ & 0.064 & 900 & $3.14\times10^{-1}$ \\
        7.5 & $1.95\times10^{-1}$ & 0.121 & 900 & $1.94\times10^{-1}$ & 0.101 & 900 & $2.03\times10^{-1}$ \\
        7.0 & $1.14\times10^{-1}$ & 0.147 & 900 & $1.14\times10^{-1}$ & 0.135 & 900 & $1.17\times10^{-1}$ \\
        6.5 & $6.31\times10^{-2}$ & 0.196 & 900 & $6.17\times10^{-2}$ & 0.191 & 900 & $6.60\times10^{-2}$ \\
        6.0 & $3.18\times10^{-2}$ & 0.262 & 900 & $3.47\times10^{-2}$ & 0.132 & 1,900 & $3.38\times10^{-2}$ \\
        5.5 & $1.59\times10^{-2}$ & 0.330 & 900 & $1.71\times10^{-2}$ & 0.202 & 900 & $1.67\times10^{-2}$ \\
        5.0 & $7.50\times10^{-3}$ & 0.445 & 900 & $8.26\times10^{-3}$ & 0.337 & 900 & $8.07\times10^{-3}$ \\
        4.5 & $3.44\times10^{-3}$ & 0.587 & 900 & $3.84\times10^{-3}$ & 0.481 & 900 & $3.80\times10^{-3}$ \\
        4.0 & $1.54\times10^{-3}$ & 0.746 & 900 & $1.73\times10^{-3}$ & 0.650 & 900 & $1.69\times10^{-3}$ \\
        3.5 & $6.75\times10^{-4}$ & 0.951 & 907 & $7.87\times10^{-4}$ & 0.818 & 900 & $8.02\times10^{-4}$ \\
        3.0 & $3.05\times10^{-4}$ & 1.154 & 900 & $3.61\times10^{-4}$ & 0.990 & 904 & $3.62\times10^{-4}$ \\ \midrule
        Sum & & & 11,807 & & & 12,804 & \\ \bottomrule
        \end{tabular}
    \end{table}
\end{document}